%% file: ms_arXiv_2.tex
\documentclass[iop,apj,numberedappendix,appendixfloats]{emulateapj}  % To get the tables correct!

\usepackage{amsmath,amsthm,amssymb}
\usepackage{bm,graphicx,color,times}
\usepackage{threeparttable} %for the table notes
\usepackage[normalem]{ulem} %%required for various underlining styles
\usepackage{enumitem} %control of the list indentation & vertical spacing between the items
\setlist{listparindent=\parindent,leftmargin=*}%indented paragraphs within lists,
\usepackage{cancel} %nice ways to cross out maths

\newcommand{\phz}{{\phantom{0}}} %empty space, 0-wide, useful to align table entries
\newcommand{\phzz}{{\phantom{00}}} %empty space, 00-wide, useful to align table entries
\newcommand{\dd}{\mathrm{d}}    %differential
\renewcommand\ion[2]{#1$\,${\footnotesize\rmfamily{#2}}\relax}
\newcommand{\crit}{_\mathrm{cr}}  %critical, subscript
\newcommand{\h}{_\mathrm{h}}  %hot, subscript
\newcommand{\SN}{_\mathrm{SN}}  %supernova, subscript
\newcommand{\B}{B}  				%total magnetic field
\newcommand{\BB}{\overline{B}}  	%mean field
\newcommand{\BBv}{\overline{\vect{B}}}  	%mean field, vector
\newcommand{\Bb}{b}  	%random field
\newcommand{\SFRs}{_\mathrm{*}}  				%star formation rate, subscript
\newcommand\vect[1]{\pmb{#1}}			%vector
\newcommand{\I}{_\mathrm{I}}  %subscript neutral hydrogen

\newcommand{\lsim}{\mathrel{\mathchoice {\vcenter{\offinterlineskip\halign{\hfil
$\displaystyle##$\hfil\cr<\cr\sim\cr}}}
{\vcenter{\offinterlineskip\halign{\hfil$\textstyle##$\hfil\cr<\cr\sim\cr}}}
{\vcenter{\offinterlineskip\halign{\hfil$\scriptstyle##$\hfil\cr<\cr\sim\cr}}}
{\vcenter{\offinterlineskip\halign{\hfil$\scriptscriptstyle##$\hfil\cr<\cr\sim\cr}}}}}

\newcommand{\erg}{\,{\rm erg}}  %erg
\newcommand{\cm}{\,{\rm cm}}    %cm
\newcommand{\G}{\,{\rm G}}      %Gauss
\newcommand{\p}{\,{\rm pc}}     %parsec
\newcommand{\km}{\,{\rm km}}    %km
\newcommand{\kms}{\km\s^{-1}}   %km/s
\newcommand{\kpc}{\,{\rm kpc}}  %kpc
\newcommand{\mkG}{\,\mu{\rm G}} %microGauss
\newcommand{\s}{\,{\rm s}}      %seconds
\newcommand{\yr}{\,{\rm yr}}    %years
\newcommand{\Gyr}{\,{\rm Gyr}}  %gigayears

\definecolor{webgreen}{rgb}{0,.5,0}
\definecolor{webbrown}{rgb}{.6,0,0}
\definecolor{purple}{rgb}{0.5,0,.5}
\shorttitle{Magnetic fields in a sample of nearby spiral galaxies}
\shortauthors{\sc Van Eck et al.}

\begin{document}

%---------------------------------------------------------------------------
\title{Magnetic Fields in a Sample of Nearby Spiral Galaxies}
\author{C.~L.\ Van Eck}
\affil{Department of Astrophysics, Faculty of Science, Radboud University Nijmegen,
PO Box 9010, 6500 GL Nijmegen, The Netherlands}
%\and
\author{J.~C.\ Brown}
\affil{Department of Physics and Astronomy, University of Calgary, Canada, T2N 1N4}
\and
\author{A.\ Shukurov and A.\ Fletcher}
\affil{School of Mathematics and Statistics, Newcastle University,\\
Newcastle upon Tyne, NE1 7RU, U.K.}
\email{c.vaneck@astro.ru.nl; jocat@ucalgary.ca; andrew.fletcher@ncl.ac.uk; anvar.shukurov@ncl.ac.uk}

%------------------------------------------------------------------------------
\begin{abstract}

Both observations and modeling of magnetic fields in the diffuse interstellar gas
of spiral galaxies are well developed but the theory has been confronted
with observations for only a handful of individual galaxies. There is now
sufficient data to consider statistical properties of galactic
magnetic fields. We have collected data from the literature on the magnetic
fields and interstellar media (ISM) of 20 spiral galaxies, and tested for
various physically motivated correlations between magnetic field and ISM parameters.
Clear correlations emerge between the total magnetic field strength and molecular
gas density as well as the star formation rate. The magnetic pitch angle exhibits 
correlations with the total gas density, the star formation rate and the strength of 
the axisymmetric component of the mean magnetic field.
The total and mean magnetic field strengths exhibit noticeable degree of
correlation, suggesting a universal behavior of the degree of order in galactic
magnetic fields. We also compare the predictions of galactic dynamo theory to observed 
magnetic field parameters and identify directions in which theory and observations might 
be usefully developed. 

\end{abstract}


%----------------------------------------------------------------------------
\section{Introduction}

Magnetic fields are recognized as an essential component of the interstellar
medium (ISM) in spiral galaxies. In particular, they confine cosmic rays \citep{Betal90},
contribute to disk-halo interactions \citep{Norman89, Kahn93},
transfer angular momentum in gas clouds to allow stars to form
\citep[e.g.][]{zh97} and provide vertical support of the interstellar gas
\citep{bc90,FS01}. The origin of galactic magnetic fields is plausibly
connected with dynamo action \citep{BBMSD96,S07} \citep[see, however,][]{K99},
but details of their structure and evolution remain insufficiently explored and understood,
either theoretically or observationally.

Our goal in this paper is to develop approaches to compare
theory and observations of galactic magnetic fields to complement detailed studies of
individual galaxies \citep{Beck12} with an exploration of galaxy samples using statistical
tools. As a first step in such an exploration, one has to identify specific combinations
of observable galactic parameters that control magnetic fields in the framework of each theory.

The number of galaxies with well-explored magnetic fields has increased
in recent years to a few dozen \citep[e.g.,][]{Beck07,Chyzy08, Fletcher11}.
It is now possible to begin exploring galactic magnetic fields, and their inter-connections
with other elements of the interstellar environment on a statistical level.
This has been done for individual galaxies \citep[e.g.,][]{Chyzy08,TBFBS13,Tetal13} and for  a
sample of dwarf irregular \citep{Chyzy11} and normal spiral galaxies \citep{HBLHBBB14},
with main emphasis on the radio--(far-)infrared and radio--star formation correlations,
but without any deep comparison with theoretical models of galactic magnetic fields. 
Comparisons of the predictions
of dynamo theory with observations were restricted to individual galaxies
\citep{RS81,RSS85,BSRS87,KRSS89,SS89,MSSBB98,RBE99,MSSBF01,MSESBS07}. It is compelling
and imperative to clarify how statistical properties of magnetic fields in a
sample of galaxies available compare with theoretical predictions. Apart from other
outcomes, such an analysis would be able to suggest the most efficient directions for
both observational and theoretical developments.

We present the data set used in Section~\ref{data},
and identify galactic parameters and their combinations relevant to interstellar magnetic fields
in Section~\ref{ACDI}. Comparison of the observational magnetic field parameters with
predictions of the mean-field dynamo theory can be found in Section~\ref{TGDM} and 
their relation to basic 
ISM parameters in Section~\ref{BCR}. Our results are put
into a broader perspective in Section~\ref{Disc} and summarized in Section~\ref{SaC}.
As part of our effort to keep the main text brief, we present additional details in appendices.

%------------------------------------------------------------------------
\section{Data}\label{data}

\input{table1}

\input{table2}

We surveyed the literature to collect relevant information for a sample of nearby spiral
galaxies;  our survey resulted in data for 20 galaxies. A list of these galaxies with some noteworthy 
parameters is included in Table~\ref{galaxybackground}.
In order to be considered for this study, the galaxy had to have a magnetic field
strength (or at least its average value) reported for some clearly-defined region (either by
stating the radial range, or a qualitative description of the area observed).
Then, we isolated the following parameters where they were available: strengths for the total and
mean (large-scale) magnetic fields, $\B$ and $\BB$ respectively, the strength of the axisymmetric 
component of the mean magnetic field $\BB_0$, the pitch angle of the mean 
magnetic field, $p_B$ (also referred to as the magnetic pitch angle),
the mass surface densities of atomic and molecular hydrogen, $\Sigma\I$ and $\Sigma_2$, respectively,
the surface density of star formation rate (SFR) $\Sigma\SFRs$, and the rotation curve, 
from which we calculated the angular velocity
$\Omega$ and rotational shear $S=r\,\dd\Omega/\dd r$.

The available estimates of the \textit{large-scale\/} magnetic field are obtained from
either the degree of polarization, assuming energy equipartition
with cosmic rays and most often neglecting any depolarization effects, and/or
from the Faraday rotation, by measuring the rotation measure from polarization angles at two or more different frequencies.   The former approach cannot distinguish between
the genuine large-scale magnetic field and an anisotropic random magnetic field
(summarily described as an ordered magnetic field),
whereas the latter yields the true large-scale magnetic field
weighted with thermal electron density. It is understandable then
that the equipartition estimates are systematically higher than those from 
Faraday rotation. \citet{SSFBLT12} show, with M33 as an example, that the
difference is consistent with the expected degree of anisotropy of the random
magnetic fields \citep[see also ][]{Ietal13}.
On the other hand, magnetic field estimates from Faraday rotation depend on the assumption
of the correlation between magnetic field strength and thermal electron density;
this plausibly introduces systematic bias into magnetic fields strengths obtained \citep{BSSW03}.

Fourteen galaxies in the sample have the average magnetic field strength published, but
not the precise averaging region. Therefore, we had to make some assumptions about the averaging
regions used. We chose to use the extent of polarized emission along the major axis of the
galaxy as the diameter over which to average; we defined the extent of polarized emission
as the location of the lowest contour of the polarized intensity map that is always included
in such publications.  Galaxies falling into this category are listed in
Table~\ref{tabledata} with asterisks.

For some of the nearest galaxies (e.g., M31, M33, M51 and M81), the strength of the 
mean
magnetic field is reported for a series of concentric rings of well-defined radii.
These estimates are more reliable as they are obtained from the differences 
between the polarization angles observed at several wavelengths and allow
for depolarization effects if necessary. 
In such
cases, the averaging region for the magnetic field is well defined, and it was possible to
ensure that the values used for the other parameters cover the same region of the galaxy.
For NGC~6946, the energy density of the azimuthally-averaged total and mean magnetic fields
were published in graphical form as continuous functions of the galactocentric radius.
For IC~342, an average magnetic field within the galactocentric radius $13.5\kpc$ is known.
These data are shown in Table~\ref{tabledata}
without the asterisk on their radial range. We also used the data for individual rings wherever
available.

\input{table3.tex}

%-----------------------------------------------------------------

For five of the galaxies in the sample (M31, M33, M51, NGC~1097 and NGC~1365) detailed modeling had been used
to separate the different azimuthal components of the large-scale magnetic field in different radial ranges 
(i.e., identifying the amplitudes $\BB_{m}$ of the 
azimuthal Fourier modes given by $\BB=\sum_{m}\BB_{m}\cos m(\theta-\beta_{m})$, where $\theta$ is the 
azimuthal position in the galaxy disc and $\beta_{m}$ the phase of the mode). For these galaxies we have also 
used the strength of the axisymmetric $m=0$ mode $\BB_0$ as this is expected to have the fastest growth rate 
and cover the largest radial range according to the standard mean-field galactic dynamo theory \citep{RSS88}.  

The linear resolution of
radio observations, shown in Table~\ref{galaxybackground}, often exceeds
the scale of the mean magnetic field, which is of order $1\kpc$. The data can be
corrected for unresolved gradients of magnetic field as described in Appendix~\ref{UGRMF},
but we found this correction insignificant in our sample.

Only a small fraction of the galaxies in our sample also have magnetic pitch angles reported,
most often as averages over specific radial ranges. For three galaxies (M94, NGC~253
and NGC~4414) the average pitch angle is given without specifying the averaging region.
For these galaxies, we defined the region of averaging as the extent of polarization, as we
did for the field strengths. The magnetic pitch angle values are given in
Table~\ref{tabledata}.

%----------------------------------------------------------------------
% FIGURE 1:
%----------------------------------------------------------------------
\begin{figure}
\epsscale{1.2}
\plotone{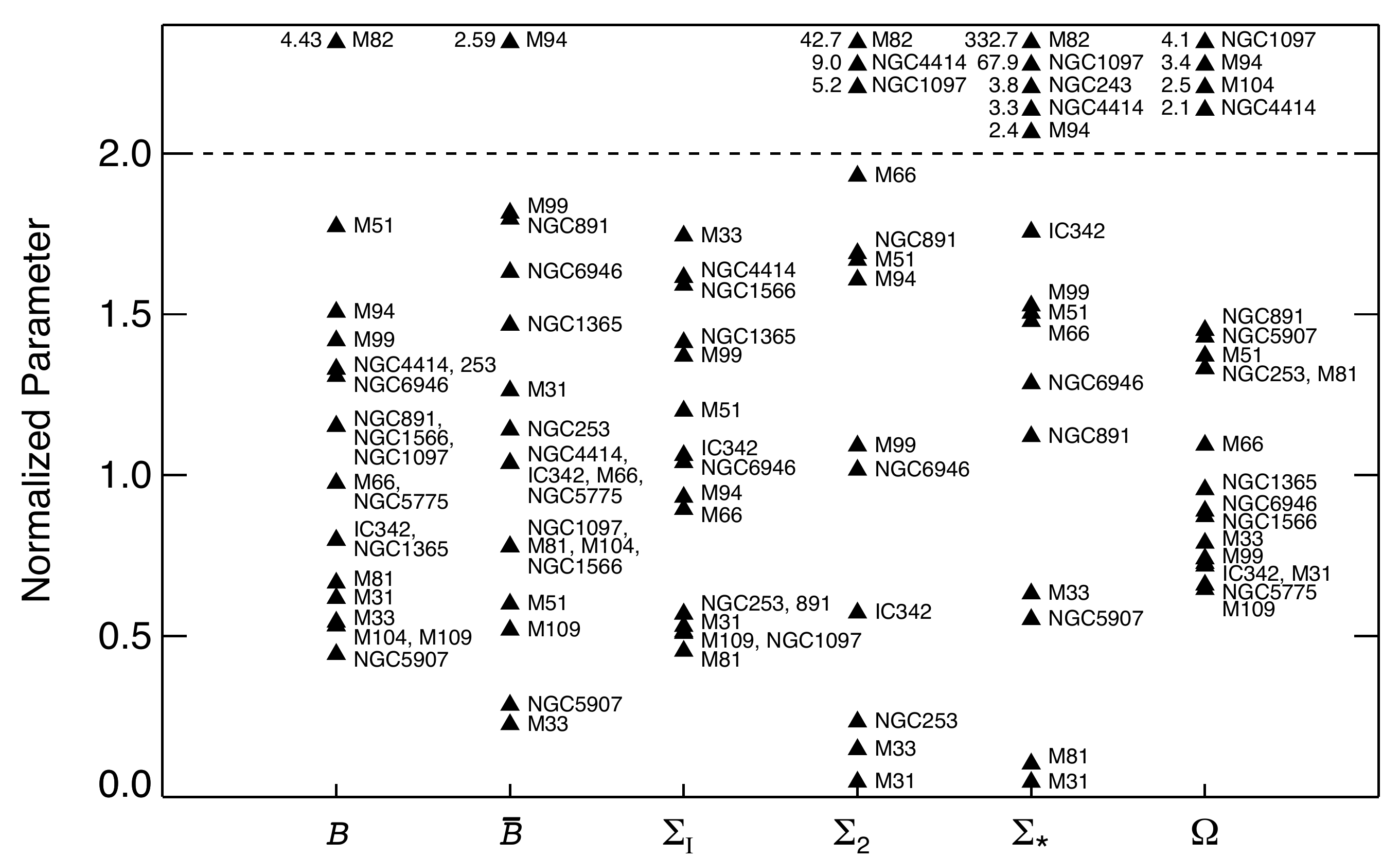}
\caption[]{\label{spread}
A visual summary of the galactic parameters in the sample, normalized to their mean values.
From left to right:
$\B$ - the total magnetic field; 
$\BB$ - the mean magnetic field ;
$\Sigma\I$ - the surface density of neutral hydrogen (\ion{H}{I});
$\Sigma_2$ - the surface density of molecular hydrogen ($H_2$); 
$\Sigma\SFRs$ - star formation rate per unit area; 
$\Omega$ -  angular velocity.   Average values for each parameter were calculated using 
galaxies below the dashed line.  As an example, the total magnetic field for M82 has 4.43 times the average total magnetic field for the remaining galaxies in the column (all below the dashed line), while M66 is roughly at the average (as indicated by its position close to 1.0 normalized parameter). From Table  \ref{tabledata}, $\B$ for M82 is 50 $\mu$G $ = 4.43\times 11.29 \; \mu$G, while $\B$ for M66 is 11 $\mu$G.  This method of normalization was done solely to avoid crowding in the plot and allow individual galaxies  to be identified in the distribution of the parameters. 

}
\end{figure}
%------------------------------------------------------------------------

For other relevant parameters of galaxies, we computed the area-averages by
integrating each quantity of interest over the same radial range where the
magnetic field was averaged, as given in Table~\ref{tabledata}. Since the 
magnetic field strength and pitch angle data were averaged over different radial ranges 
in some cases, we computed the averages separately for each such radial range.
The average value of a product of variables can be rather different from
the product of their averages; the mean values shown and used in such cases were obtained by
averaging the corresponding products.  The sources of the data are given in
Table~\ref{tableBrefs1}.

The data we collected are summarized in Figure~\ref{spread} which shows some parameters normalized
to their mean value in the sample. This figure  illustrates the properties of our galaxy sample. 
The magnetic pitch angle has the narrowest distribution in the sample,
$8^\circ\leq|p_B|\leq37^\circ$.
The magnetic field strength has a relatively narrow distribution, 
with a spread of one order of magnitude, $5\lsim\B\lsim50\mkG$ and
$1\lsim\BB\lsim10\mkG$.
However, a more physically relevant quantity, the magnetic energy density,
has a spread of two orders of magnitude, comparable to that in the other parameters.
For comparison, the 
mass 
surface densities of molecular
hydrogen and the star formation rate 
also 
span two orders of magnitude,
$0.5\lsim\Sigma_2\lsim100\,M_\odot\p^{-2}$ and
$0.4\lsim\Sigma\SFRs\lsim30\,M_\odot\p^{-2}\Gyr^{-1}$
(excluding M82 that has an extremely high 
surface density of star formation rate
as well as other parameters because the data refer to the central part of the galaxy). 
Altogether, the sample contains 20 galaxies in almost
all the variables required for analysis based on the concepts presented below.

Before verifying the theoretical predictions discussed in Section~\ref{ACDI}, we perform
a consistency check of the data collected, and their averaging, by testing the
Schmidt--Kennicutt law \citep{Kennicutt89}, an empirical relation between the
surface mass density of the interstellar gas $\Sigma=\Sigma_\mathrm{I}+\Sigma_2$
and that of star formation $\Sigma\SFRs$. Indeed, the data are consistent with
\[
\Sigma\SFRs \propto \Sigma^{1.25\pm 0.41}\,,
\]
with a Spearman correlation coefficient of $0.52$. This agrees fairly well with
\citet{Kennicutt89}, who found a power law index of $1.3\pm 0.3$, with
Spearman correlation coefficients between $0.49$ and $0.76$.

We cannot be similarly encouraged by the quality of the data on the 
galactic magnetic fields. Most of the estimates of both the total and mean magnetic field strengths
are determined from energy equipartition with cosmic rays, often using uncertain
estimates of the path length and the relativistic proton-electron ratio, imperfect (if any) 
separation of thermal and nonthermal contributions to the total radio intensity, etc. 
Furthermore, the size of the averaging region used in the estimates varies 
vastly between the galaxies, and it is not clear in advance if such data can be used to
test any theory at all. Therefore, we first focus on a few nearby galaxies where parameters of the
mean magnetic field were obtained from well-defined fits of azimuthal Fourier modes 
to multi-frequency observations of the polarization angles within
relatively narrow rings that pass statistical goodness-of-fit tests. These are the
best data on the galactic mean magnetic fields available. Our discussion of the magnetic fields
obtained from equipartition arguments is only exploratory and serves mostly to assess their quality and
clarify the most important improvements required in the interpretation of the radio astronomical data.

%----------------------------------------------------------------------
\section{Astrophysical context for the data interpretation}\label{ACDI}

Compression and stretching are the two fundamental processes that can 
affect the magnetic field, $\vect{\B}$. In a random flow, they can result in the 
generation of a self-sustained magnetic
field via various forms of dynamo action, both at ``large'' scales (larger than the
correlation scale of the random flow) to produce a mean (large-scale) magnetic field
$\BBv$, and at the range of scales of the random motions themselves (the ``small'' scales) to
generate
a random magnetic field $\vect{\Bb}$ in a wide range of scales. Interstellar turbulence
is transonic, so that both stretching and compression affect interstellar magnetic fields.
In this section we briefly discuss
the most important processes thought to affect interstellar magnetic fields
and their dependence on galactic parameters.

%----------------------------------------------------------------------
\subsection{Compression and stretching of magnetic fields}\label{CSMF}

In relatively simple flows, connection between gas density and velocity resulting from the
continuity equation can be tractable, and magnetic field strength can be expressed as a
function of gas density $\rho$ alone.
For example, magnetic flux through any moving contour is conserved in an ideal plasma. 
Together with mass conservation, this results in a power-law
dependence of magnetic field strength on gas density $\rho$,
\begin{equation}\label{rhon}
\B\propto\rho^k\,,
\end{equation}
with $k=2/3$ for a spherically symmetric collapse, $k=1/2$ for an anisotropic
compression into a flattened cloud via a sequence of quasi-equilibrium
states \citep{MP84}, and $k=1$ for a one-dimensional compression, as in a shock.

Correlations consistent with Eq.~\eqref{rhon}, with $k\approx0.65$ in denser clouds with particle
number density $n$ in excess of $300\cm^{-3}$ \citep{Calzetti10}, are well known from
Zeeman measurements \citep{Troland86}.
However, 
\citet{B00} noticed that magnetic field strength in dense interstellar clouds
has a tighter correlation with kinetic energy density within the cloud,
\begin{equation}\label{rhov}
\B^2\propto\rho v^2\,,
\end{equation}
rather than the gas density alone. Scaling of magnetic field strength with kinetic energy
density is a typical feature of dynamo mechanisms, a kinetic-to-magnetic energy conversion.

For our study, the physically distinct correlations as presented in 
\eqref{rhon} and \eqref{rhov} cannot be distinguished between. The intra-cloud 
velocity dispersions are not available for external galaxies
as the resolution of either CO or \ion{H}{I} observations is not sufficient
to separate the intra-cloud velocity dispersion from that arising from the relative 
random motions of individual clouds.

Compression and stretching at large scales introduce anisotropy to an otherwise
isotropic random magnetic fields \citep{Beck05}. Synchrotron emission from an
anisotropic random magnetic field can be polarized
\citep{SBSBBP98}, making it more
difficult to interpret the polarization of galactic radio emission in terms of large-scale
magnetic fields. The anisotropy expected from stretching by the galactic
differential rotation can account for a degree of polarization of order 10\%
\citep[Section 2.1 in][]{SSFBLT12}. Calculations of anisotropic random magnetic fields
produced by both galactic differential rotation and large-scale shocks can be found
in \citet{Beck05}.

Compression by spiral arms and stretching by the associated streaming motions
affect systematically the orientation of magnetic field at both large and small scales.
In particular, magnetic field lines are refracted to a better alignment with the
spiral arms within them. If the angle between the arm axis and magnetic field is
$p_1$ between the arms and $p_2$ within them, one-dimensional compression leads to
\begin{equation}\label{p_arms}
\tan p_2=\frac{\rho_1}{\rho_2}\tan{p_1} <\tan{p_1}\,,
\end{equation}
where $\rho_1$ and $\rho_2$ are the gas densities between the arms and within the arm,
respectively. As an example, $p_2\approx5^\circ$ for $p_1=20^\circ$ and $\rho_2/\rho_1=4$.
Streaming motions can improve the alignment even further.

%-----------------------------------------------------------------------------
\subsection{Origin of galactic magnetic fields}\label{GDSSMF}

Large-scale (mean) magnetic fields coherent at a scale of order 1--$10\kpc$
are a common feature of spiral galaxies; they have been detected in all spiral
galaxies observed with adequate sensitivity and resolution. Their widespread
presence calls for a universal mechanism of their generation and maintenance.
Two such mechanisms have been suggested: the turbulent mean-field dynamo theory
\citep{RSS88} and the primordial field theory \citep{K99}. Unlike the dynamo
theory, the primordial concept does not lead to any specific predictions for the
galactic magnetic fields, and its main unresolved problem is to explain consistently
such basic parameters as the magnetic pitch angle and the predominantly
quadrupolar parity of the large-scale magnetic fields \citep{S07}.

A primordial magnetic field may still serve as a seed magnetic field for galactic dynamos
\citep{KZ08}, but otherwise this theory appears to be ruled out by the current state of
observations. The cosmic dynamo theory does not require any hypothetical
primordial magnetic field to launch the large-scale dynamo action. Instead, magnetic fields
produced by battery mechanisms in stars, expelled into the ISM by stellar winds
and supernova explosions, and then further amplified at a time scale of order
$10^7$--$10^8\yr$ by the fluctuation dynamo in the interstellar medium
\citep{S07,SS08} provide a seed magnetic field whose effective strength is of order
$10^{-9}\G$ at a scale of a few kiloparsecs \citep{Ruzmaikin88,BPSS94}, which is quite
sufficient to explain the observed large-scale magnetic fields in the framework of the
mean-field dynamo theory.

Galactic dynamo theory is well developed in the \textit{kinematic\/} regime that
describes the early stages of magnetic field growth, when the Lorentz force is
still negligible  in comparison with other forces in the ISM. However, the
non-linear behavior of the interstellar magnetic field, when it approaches a
statistically steady state, remains controversial,
for both large- and small-scale dynamos.

In what follows, we present estimates of the steady-state strength of the 
large-scale
magnetic field in spiral galaxies $\BB$ that follow from the mean-field dynamo theory,
with allowance for the uncertainties in its non-linear aspects.
Throughout this paper, we use cylindrical polar coordinates $(r,\phi,z)$ with the
origin at the galaxy's center and the $z$-axis aligned with the galactic angular
velocity $\vect{\Omega}$.

%----------------------------------------------------------------------
\subsection{Basics of galactic dynamos}\label{BGD}

Star formation in spiral galaxies, resulting in supernova explosions and
galactic fountains and winds, drives ubiquitous transonic and supersonic
random flows at scales of order $l\simeq100\p$ and less. Together with the
velocity field, interstellar magnetic fields can be naturally and usefully
represented as a sum of two physically distinct parts, a mean field $\BBv$
(at scales of order 1\,kpc and more) and a random field $\vect{\Bb}$ (at scales
100\,pc and less). The mean and random magnetic fields are produced by
different (albeit related) physical mechanisms. As we discuss below, the former
are produced by the mean-field dynamo action which relies on density stratification
of the galactic discs and their differential rotation. At kiloparsec scales, magnetic
fields are further modified by the spiral pattern and galactic outflows.

Like other constituents of the ISM, the large-scale magnetic field is strongly
affected by the multi-phase gas structure, and different phases play different roles in 
its generation and evolution \citep{S07}. Dense, cold clouds occupy a negligible
fraction of the total volume, and field lines in the densest molecular clouds are subjected
to enhanced magnetic reconnection, so this phase is not likely to host the large-scale
dynamo. The hot gas is buoyant and leaves the disk for the galactic halo on a time 
scale shorter than the mean-field dynamo amplification time. The only pervasive, diffuse 
phase left as a candidate to host the mean-field dynamo action is the warm gas,
which remains in a well-defined layer (despite being partially entrained in galactic 
outflows), and is partially ionized. The warm gas occupies a significant fraction of the volume, 
hence probably forms a connected region, and would thereby be able to accommodate the magnetic field 
coherent over kiloparsec scales. Thus, the warm phase has all the properties required 
to be the site of the mean-field dynamo action. Its parameters will be used in our 
discussion of the large-scale magnetic fields.

Random magnetic fields can be produced by another dynamo mechanism, the
fluctuation dynamo, which acts, to a large extent, independently of the amplification 
of the large-scale magnetic field. This mechanism does not require anything more
than a random plasma flow of sufficient intensity. Random magnetic fields are also 
produced by tangling of the mean magnetic field (in fact, this process is an 
essential part of the mean-field dynamo) and further modified by interstellar shocks. 
Hence, the mean and random magnetic fields are sensitive to distinct features of 
the plasma flow and thus depend on different galactic parameters. Therefore, they 
should be carefully separated before any meaningful 
relations to galactic parameters can be established.

The intensity of the induction effects producing a large-scale magnetic field, 
relative to its dissipation by Ohmic resistivity, which is enhanced by the
tangling of magnetic field lines by the random flow, can be quantified using 
the (dimensionless) turbulent magnetic Reynolds numbers
\begin{equation}\label{RaRo}
R_\alpha=\frac{\alpha h}{\beta}\,,
\quad
R_\omega=\frac{S h^2}{\beta}\,,
\end{equation}
where
\begin{equation}\label{alp}
\alpha\simeq l^2\Omega/h
\end{equation}
is a measure of the large-scale induction effects due to the 
helical random flows 
(arising from the 
systematic effects of the Coriolis force on the stratified galactic turbulence), 
\begin{equation}\label{beta}
\beta\simeq
\tfrac13 lv\,,
\end{equation}
is the turbulent magnetic diffusivity, $h$ is the
pressure scale height, $l$ and $v$ are the turbulent scale length and velocity, and
$S=r\dd\Omega/\dd r$ is the large-scale velocity shear rate due to the
galactic differential rotation. In the widely used approximation of an 
$\alpha\omega$-dynamo
(where the induction effects of the galactic differential rotation are considered
to be much stronger than the production of the large-scale magnetic field by
galactic turbulence), 
it is useful to introduce the product of the two
Reynolds numbers known as the dynamo number,
\begin{equation}\label{Dyn}
D=R_\alpha R_\omega\,.
\end{equation}
The dimensionless dynamo control parameters $R_\alpha$, $R_\omega$ and $D$
vary with the galactocentric radius $r$, mainly because $\alpha$, $\Omega$,
$S$ and $h$ depend on $r$. All the variables entering these definitions are 
observable, at least in principle:
\begin{align}\label{Raln}
R_\alpha\simeq&
3
\frac{l\Omega}{v} 
= 0.75
		\left(\frac{\Omega}{25\km\s^{-1}\kpc^{-1}}\right)\nonumber\\
&\times
		\left(\frac{l}{0.1\kpc}\right)
 		\left(\frac{v}{10\kms}\right)^{-1},
\end{align}
and
\begin{align}\label{Romn}
R_\omega\simeq&
3
\frac{Sh^2}{lv}
=19
		\left(\frac{S}{25\km\s^{-1}\kpc^{-1}}\right)
		\left(\frac{h}{0.5\kpc}\right)^2\nonumber\\
&\times
		\left(\frac{l}{0.1\kpc}\right)^{-1}
 		\left(\frac{v}{10\kms}\right)^{-1},
\end{align}
where $h$ is the (pressure) scale height of the warm gas, as above, and 
$r$ is the galactocentric radius. Since $\Omega$ usually decreases with $r$,
we have $S<0$,  $D<0$ and $R_\omega<0$ in most cases.

Thus, the dynamo number becomes:
\begin{align}\label{Dpr}
D\simeq&
9
\frac{\Omega Sh^2}{v^2}\nonumber\\
=&14
		\left(\frac{\Omega S}{625\km^2\s^{-2}\kpc^{-2}}\right)
		\left(\frac{h}{0.5\kpc}\right)^2
 		\left(\frac{v}{10\kms}\right)^{-2}.
\end{align}

In most (if not all) spiral galaxies, the mean-field dynamos generate a basic axisymmetric 
mean magnetic field which is further modified by the spiral pattern, bar, etc., to add 
non-axisymmetric components \citep{BBMSD96}. Therefore, predictions of the dynamo theory
should be compared not with the total mean magnetic field observed, but with its
axisymmetric part. We have done this whenever possible (see Table~\ref{tabledata}),
but most observational results do not provide this information, limiting the number of galaxies
we could compare with.

%--------------------------------------------------------------
\subsection{Mean magnetic field strength}\label{RMFS}
The mean-field dynamo theory offers a range of estimates of the steady-state
strength of the mean magnetic field whose complexity reflects the amount of
detail in the underlying theory. We present here three such estimates: first a 
generic case, which only relies on the fundamental aspects of dynamo
action, and then two further, more involved, models.

%--------------------------------------------------------------
\subsubsection{Equipartition of magnetic and turbulent energies}\label{GA}
The simplest estimate of magnetic field strength in the steady state relies on
the fact that magnetic field energy is obtained from the kinetic energy of
interstellar turbulence, so the statistically steady state can be expected to
have comparable magnetic and turbulent kinetic energies, 
$\BB^2=4\pi\xi \rho v^2$, where $\rho$ is the gas density, 
$v$ is the root-mean-square random velocity, and $\xi$ is a factor of order unity. 
Later refinements of this estimate in Sections~\ref{MagBa} and \ref{MHB} 
represent, in fact, a clarification of the dependence of $\xi$ on physical parameters.

Under the simplest bifurcation (which takes place in the
mean-field dynamo in a thin galactic disc), the 
energy density of the steady-state large-scale
magnetic field will also be proportional to the deviation of the dynamo
control parameter (i.e., the dynamo number) $D$ from its marginal (critical) value $D\crit$
with respect to the dynamo action (such that the large-scale magnetic field 
grows if $|D|>|D\crit|$ and decays otherwise):
\begin{equation}\label{eqkin}
\BB^2\simeq 4\pi\rho v^2 \left(\frac{D}{D\crit}-1\right)
\text{ for } \frac{D}{D\crit}>1\,,
\quad
\BB=0 \text{ otherwise,}
\\
\end{equation}
where $\rho$ is the density of the warm gas and we use $\xi=1$. 

Equation \eqref{eqkin} can be written in terms of the surface density
of the warm gas, approximated by that of neutral hydrogen $\Sigma\I=2h\rho$:
\begin{align}\label{energy_bal_obs}
\BB^2=&(0.8\mkG)^2 
		\left( \frac{D}{D\crit}-1\right)
		\left(\frac{\Sigma\I}{1\,M_\odot\p^{-2}}\right)
		\left(\frac{h}{0.5\kpc}\right)^{-1} \nonumber\\
&\times
		\left(\frac{v}{10\kms}\right)^2
\end{align}
for $D/D\crit>1$.
This expression establishes the relation between the strength of the mean magnetic field
and other directly observable galactic parameters resulting from the general
concept of equipartition between magnetic and turbulent energy densities. 
It will be tested for the sample galaxies in Sections~\ref{TGDM} and \ref{BCR}.

In the simplest dynamo models, $D\crit$ is a constant depending on the 
specific form of $\alpha$ as a function of $z$, of which we only know 
that $\alpha$ is an odd function of $z$ and presumably varies with $r$ as 
given in Eq.~\eqref{alp}. For a quadrupolar magnetic field, predominant 
in a thin-disk dynamo, the critical dynamo number remains within a relatively
narrow range, $-4\leq D\crit\leq-13$, for very broad and diverse range of the 
model forms of $\alpha(z)$ \citep{RSS88}. When using such simple dynamo models,
we select $D\crit=-8$ as a suitable estimate near the middle of this range; 
this corresponds to $\alpha\propto\sin(\pi z/h)$.

We discuss in Section~\ref{MHB}, however, that $D\crit$ may depend on the
speed of the galactic outflow (fountain or wind); then Eq.~\eqref{DcrU} is 
appropriate.

%--------------------------------------------
\subsubsection{Magnetostrophic balance}\label{MagBa}

A more physically detailed estimate of the steady-state mean magnetic field can be
obtained by considering more carefully the mechanism by which the dynamo may saturate. 
The generation of the large-scale magnetic field relies on the mean helicity of interstellar 
random flows; the mean helicity is due to density stratification and galactic rotation 
combining to twist rising or sinking turbulent cells via the azimuthal component
of the Coriolis force $\mathcal{C}=2\rho[\vect{v}\times\vect{\Omega}]_\phi\simeq2\rho
v_r\Omega$, written in the local cylindrical frame centered at the expanding turbulent
cell. With the 
$z$-axis aligned with the galactic
angular velocity $\vect{\Omega}$,
the radial (expansion) velocity $v_r$ within the cell follows from the mass conservation
$\nabla\cdot\vect{v}=0$ as $v_r\simeq v_z l/h$ in terms of the vertical 
component $v_z$. The azimuthal component of the Lorentz force produced by
the large-scale magnetic field perturbed at the turbulent scale $l$ is given by 
$\mathcal{L}=(4\pi)^{-1}[(\nabla\times\vect{B})\times\vect{B}]_\phi\simeq B_rB_\phi/(4\pi
l)$. The steady-state strength of the large-scale magnetic field then follows from the 
balance of the Coriolis and Lorentz forces, $\mathcal{C}+\mathcal{L}\simeq0$ 
(the magnetostrophic balance), as \citep[see also ][]{Ruzmaikin88}
\[
\BB_r\BB_\phi\simeq-\frac{8\pi}{\sqrt3}\rho v\alpha\,,
\]
where we have used Eq.~\eqref{alp} and assumed isotropy of the interstellar turbulence, 
$v_z=v/\sqrt3$. This estimate relies on the plausible assumption that the dynamo 
action settles to a steady state because the Lorentz force affects the turbulent flow by 
opposing the Coriolis force that makes the flow helical, i.e., the back-reaction of
magnetic field on the flow affects primarily the flow helicity ($R_\alpha$) rather than the 
differential rotation ($R_\omega$) which is supported by the stronger gravitational forces.

The radial and azimuthal components of the large-scale magnetic fields $\BB$ are related
via the pitch angle of magnetic lines,
\[
\BB_r=\BB\sin p_B\,,
\qquad
\BB_\phi=\BB\cos p_B\,,
\]
and $p_B$ can be taken either from observations or from theory; various theoretical
estimates of $\tan p_B$ can be found in Section~\ref{MPA}.

In terms of observable parameters, and including the same factor with 
$D/D\crit-1$ as above, we obtain
\begin{align}\label{strophic_bal_obs}
\BB^2
\simeq&	
{-}
\frac{16\pi}{\sqrt3}\,\frac{\rho v\alpha}{\sin2p_B}	\left( \frac{D}{D\crit}-1\right)\nonumber\\
=&
{-}
\frac{(0.3\mkG)^2}{\sin2p_B}	\left( \frac{D}{D\crit}-1\right)
		\left(\frac{\Sigma\I}{1\,M_\odot\p^{-2}}\right)	
		\left(\frac{l}{0.1\kpc}\right)^2\nonumber\\ 
&\times
   	\left(\frac{h}{0.5\kpc}\right)^{-2}    
		\left(\frac{v}{10\kms}\right)
		\left(\frac{\Omega}{25\kms\kpc^{-1}}\right),
\end{align}
and note that $\BB^2>0$ as long as $p_B<0$  (and $D/D\crit>1$).

%-------------------------------------------------------------------------
\subsubsection{Magnetic helicity balance}\label{MHB}

The mean-field dynamo action 
 can 
be saturated not via the magnetostrophic balance, but rather because the dynamo 
action is suppressed by the build-up of magnetic helicity in the galactic 
disk before the balance is achieved. \citet{SSSB06} showed that galactic outflows
(fountains or winds) can prevent a catastrophic quenching of the dynamo action
by the accumulation of the magnetic helicity, and \citet{SSS07} provide an estimate 
of the resulting steady-state mean magnetic field. 
\citet{KMRS00} suggested 
that the turbulent diffusion of magnetic helicity can provide another way to remove it 
from the dynamo region \citep[see also][]{KMRS02,KMRS03}. Although 
the diffusive helicity transport still 
needs to be justified rigorously, we also include it in our estimates 
following \citet{CSS14}. There are other helicity fluxes that may contribute to the
nonlinear state of the mean-field dynamo \citep{VC01,VS14}, but they 
are less certain, depend on the anisotropy
of the random flow, and are thus harder to quantify in terms of observable quantities.
As we neglect them, it can be expected that the the steady-state magnetic field strength 
is somewhat underestimated, and therefore we focus on its dependence on galactic parameters 
rather than its magnitude. 

The resulting estimate of the mean magnetic field strength, that allows for
both advective and diffusive fluxes of magnetic helicity, is based on a fully
nonlinear theory, and thus the factor $D/D\crit-1$ emerges automatically.
As shown by \citet{CSS14} this estimate has the form
\begin{equation}\label{hel}
\BB^2\simeq  \left(\frac{l}{h}\right)^2
\frac{2\pi\rho v^2 (R_U+\pi^2)}{1-3\sqrt{2}\cos^2(p_B)/8}\, \left(\frac{D}{D\crit}-1 \right)\,,
\end{equation}
where $R_U$ is the turbulent Reynolds number for $U_z$, the mass-averaged vertical 
velocity in the disk,
\begin{equation}\label{Ru}
R_U=\frac{U_z h}{\beta}\,,
\quad
\mbox{with}
\quad
U_z\simeq f V_z \frac{n\h}{n}\,,
\end{equation}
 $f$ is the fraction of the disk surface occupied by the outflow,
mostly the hot gas, $V_z$ is the bulk vertical velocity of the hot
gas, and $n\h\simeq10^{-3}\cm^{-3}$ and $n\simeq0.1\cm^{-3}$ are the gas
number densities in the hot and warm phases, respectively (assuming that the
warm gas hosts the mean magnetic field, whereas the outflow mainly 
consists of the hot gas). With 
$f=0.1$, the fractional area of the disc surface occupied by the
OB associations and
chimneys \citep{Norman89} and $V_z\simeq200\kms$, typical of galactic fountains, 
this yields $U_z\simeq0.2\kms$.

The gas outflow also hinders the dynamo action, and thus affects the critical dynamo number
that in fact depends on the outflow speed \citep{CSS14}: 
\begin{equation}\label{DcrU}
D\crit\simeq-\left(\frac{\pi}{2}\right)^5\left(1+\frac{R_U}{\pi^2}\right)^2\,, 
\end{equation}
so that $|D\crit|$ increases with $R_U$. 
The critical dynamo number obtained for $R_U=0$ in this approximation is 
$D\crit=-(\pi/2)^5\approx-9.6$, rather than $D\crit=-8$ as adopted above. 
We neglect this difference in view of the approximate nature of the 
solutions and parameter values used.

Estimates of $V_z$ and $U_z$
are uncertain as they involve complex connections between star formation,
the multi-phase structure of the ISM and the physical mechanisms of launching an outflow.
Two plausible estimates given in Appendix~\ref{OS} yield
consistent results,
\begin{align}\label{RUn}
R_U&\simeq0.45
		\left(\frac{U_z}{0.3\kms}\right)
		\left(\frac{h}{0.5\kpc}\right)\nonumber\\
&\times
		\left(\frac{l}{0.1\kpc}\right)^{-1}
		\left(\frac{v}{10\kms}\right)^{-1}\,,
\end{align}
with $V_z$ given by Eq.~\eqref{VzE} or Eq.~\eqref{VzML}.

Equation~\eqref{hel} for the steady-state mean magnetic field 
can be expressed in terms of directly observable quantities using
Eqs~\eqref{Dyn}--\eqref{Romn} for $D$ and $R_\omega$,  Eq.~\eqref{DcrU} for $D\crit$
and Eq.~\eqref{RUn} for $R_U$. 
The value of $p_B$ can be taken either from observations or from the accompanying
estimate \eqref{pitch_z}. For the calculations in Section~\ref{TGDM}, we use \eqref{pitch_z},
but also tested using the observed values and found the results to be consistent
(since the effect of $p_B$ on the steady-state magnetic field is weak as long
as $|p_B|$ is sufficiently small).

%--------------------------------------------------------------------------
\subsubsection{Dynamo saturation mechanism versus star formation rate}\label{D-SFR}

The magnetic field strength established through the magnetic helicity 
balance is expected to be lower than that arising from magnetostrophic balance: 
Eq.~\eqref{hel} yields a lower value of $\BB$ than Eq.~\eqref{strophic_bal_obs} 
provided
\[
\frac{U_z}{\Omega h}\lsim 0.3\,.
\]
For a flat rotation curve ($S=-\Omega$), $U_z=1\kms$,
$\Omega=25\kms\kpc^{-1}$ and $h=0.5\kpc$, the left-hand side of this inequality
is about $0.1$.

%---------------------------------------------------------------------------
\subsection{Magnetic pitch angle}\label{MPA}
The pitch angle of the mean magnetic field $p_B$, is defined as the acute angle 
between the magnetic field direction and the tangent to the local circumference,
\[
\tan p_B=\frac{\BB_r}{\BB_\phi}.
\]
In a trailing spiral, $p_B<0$, in contrast to $p_B>0$ for a leading one. This is a readily 
observable quantity which can be used to understand the mechanism that produces
the mean magnetic field, as first suggested by \citet{KRSS89}. In particular,
the pitch angle involves the ratio of magnetic field components, and thus depends
on fewer galactic parameters than the strength of the mean magnetic field.

%-------------------------------------------
\subsubsection{Kinematic dynamo}\label{pKD}
\citet{KRSS89} showed that, in the kinematic mean-field dynamo, 
\begin{equation}\label{pitch_MB}
\tan p_B\simeq-\left(\frac{R_\alpha}{|R_\omega|}\right)^{1/2}.
\end{equation}
Using Eqs~(\ref{RaRo}) and (\ref{alp}), we obtain
\begin{equation}\label{pitch_kinematic_obs}%\label{pmlh}
\tan p_B\simeq - \frac{l}{h}\left(\frac{\Omega}{|S|}\right)^{1/2}.
\end{equation}
We have $\Omega/S=-1$ for a flat rotation curve, and $l$ presumably 
varies little with the galactocentric radius $r$. Then the magnetic pitch 
angle varies with $r$ mostly because of the disk flaring, i.e., the 
increase in $h$ with $r$.
In a normalized form, we have
\begin{equation}\label{pBkn}
\tan p_B \simeq -0.2
\left(\frac{\Omega}{|S|}\right)
\left(\frac{l}{0.1\kpc}\right)
\left(\frac{h}{0.5\kpc}\right)^{-1}.
\end{equation}

%--------------------------------------------------
\subsubsection{Magnetostrophic balance}\label{pMB}
The value of $p_B$ that arises from 
 the 
magnetostrophic balance can be estimated 
heuristically by replacing the dynamo number by its critical value,
$R_\alpha R_\omega=D\crit$.  Eq.~(\ref{pitch_MB}) then reduces to
\begin{equation}\label{pitch_MBs}
\tan p_B\simeq \frac{|D\crit|^{1/2}}{R_\omega}\,,
\end{equation}
or
\begin{equation}\label{pitch_strophic_obs}
\tan p_B\simeq -0.15
\left(\frac{R_\omega}{-19}\right)^{-1}\,,
\end{equation}
where $D\crit=-8$.

%--------------------------------------------------
\subsubsection{Magnetic helicity balance}\label{pMHB}
Under magnetic helicity balance, 
$\tan p_B\simeq (R_U+\pi^2)/{4R_\omega}=(-2D\crit/\pi)^{1/2}/R_\omega$
\citep{CSS14}, similar to Eq.~\eqref{pitch_MBs} 
but with $D\crit$ from Eq.~\eqref{DcrU}:
\begin{equation}\label{pitch_z}
\tan p_B\simeq -0.13
\left(\frac{D\crit}{-10}\right)^{1/2}
\left(\frac{R_\omega}{-19}\right)^{-1}\,.
\end{equation}

In fact, this estimate is independent of the nature of the dynamo
nonlinearity as it follows from the steady-state equations for $\BB_r$ and
$\BB_\phi$ alone \citep[see][for details]{CSS14} and applies whether or not the galactic outflow
is responsible for the dynamo saturation.
Equation~\eqref{pitch_z} can readily be expressed in terms 
of directly observable quantities using Eq.~\eqref{DcrU} for $D\crit$,
Eq.~\eqref{RUn} for $R_U$, and Eq.~\eqref{Romn} for $R_\omega$.

\input{table4.tex}

%--------------------------------------------------------------------------
\section{Testing galactic dynamo models}\label{TGDM}

The mean-field galactic dynamo theory, briefly reviewed in 
Sections~\ref{BGD}--\ref{MPA}, predicts specific dependencies of the strength
and pitch angle of the mean magnetic field on galactic parameters, which can 
be tested using observations. Such testing is the subject of this section. 
Unfortunately, we are not aware of any suitable specific predictions from the primordial 
or any other alternative theory for the origin of mean magnetic fields.

%-----------------------------------------------------------------
\subsection{Strength of the mean magnetic field}\label{dB}

For each of the three relationships for the mean magnetic field
strength, Eqs~\eqref{energy_bal_obs}, \eqref{strophic_bal_obs}, and
\eqref{hel},
we calculated  for each galaxy the predicted value of the axisymmetric mean 
magnetic field strength $\BB_0$ averaged over the same area as the observational data.  
This was only possible for galaxies where all of the data was available for the radial
ranges used. All of the dynamo models use the rotation curve; the energy-equipartition 
and magnetostrophic balance models also include the gas density, and the helicity 
balance model involves the star formation rate.

%----------------------------------------------------------------------
% FIGURE 2:
%----------------------------------------------------------------------

\begin{figure}[ht]
\epsscale{.75}
\plotone{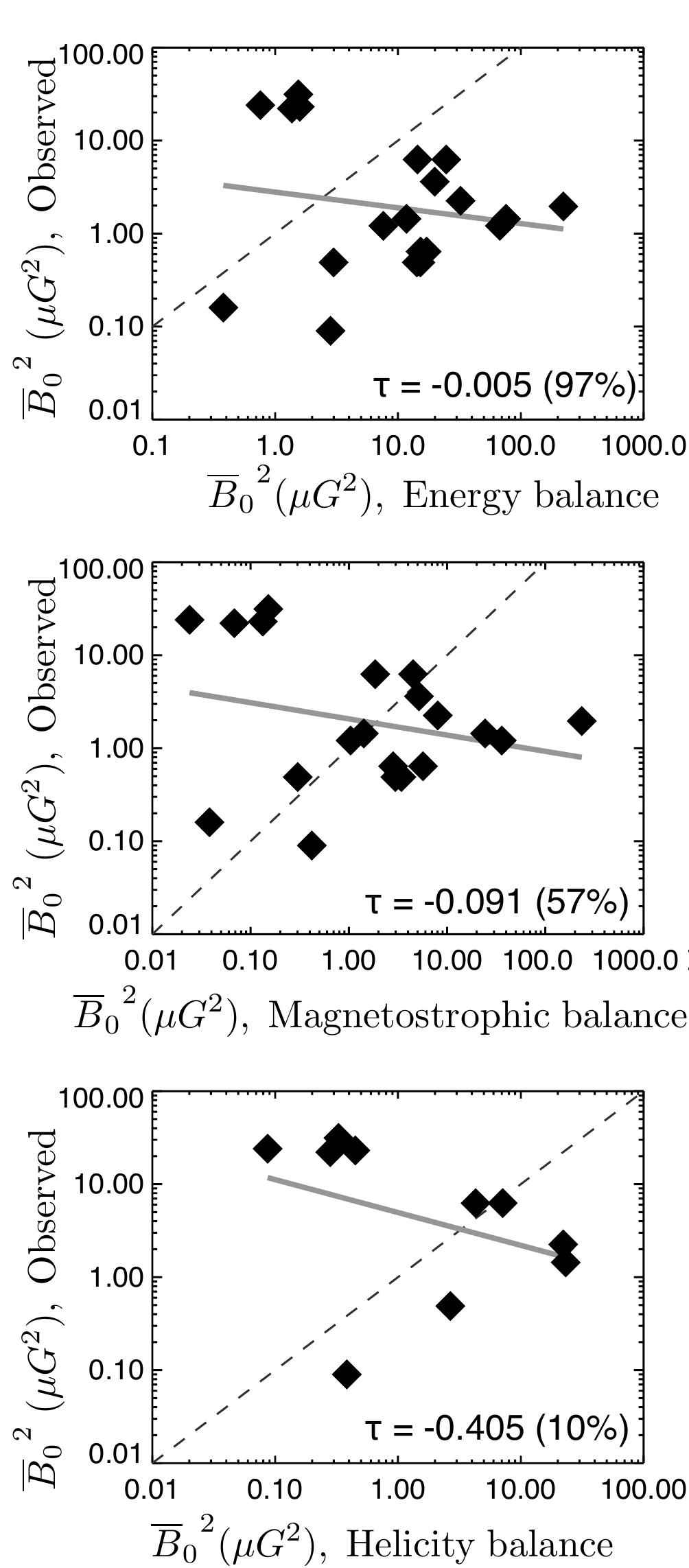}
\caption{\label{figuredynamostrength}
The observed axisymmetric mean magnetic field strength versus that from various
nonlinear dynamo models: energy equipartition with turbulence (top),
magnetostrophic balance (middle), and magnetic helicity balance (bottom).  
The gray lines are the  best-fit linear relationship; Pearson's correlation coefficient
$r$ and the significance level of the correlation (\%) are shown in each panel. 
}
\end{figure}

%%----------------------------------------------------------------------------------

To evaluate the quality of each model, we used Pearson's product-moment
correlation coefficient, with the expectation that a perfect model would
produce a correlation coefficient of unity and poor models would produce low 
correlation coefficients.  Table~\ref{tablestrengthcorrelations}
contains the calculated correlation coefficients between the observed
mean magnetic field strengths and the model predictions, with the
corresponding scatter plots shown in Figure~\ref{figuredynamostrength}.
 
The correlation coefficients for all three models are equally low, with
high significance levels.   
Thus, we can only conclude that, with the amount of data at our
disposal, none of the three models can be excluded or appears any better than
the others to any significant degree. 

%%----------------------------------------------------------------------------------

The magnetic field strength obtained from any dynamo saturation model
depends on the ratio $h^2/v^2$, treated above as a constant. However, unlike the
turbulent velocity, $h$ can vary widely between the galaxies and with galactocentric
radius within each galaxy. To assess the
consequences of treating this ratio as a constant, we varied the value of 
$h^2/v^2$, increasing and decreasing it by up to a factor of five for all the 
galaxies simultaneously, and then repeated the correlation analysis.  
We found that the correlation coefficients were changed by a minimal amount, and
our conclusions regarding all three of the dynamo models remain unaffected.

%----------------------------------------------------------------------------------

%------------------------------------------------------------------------------------
\subsection{Pitch angle of the mean magnetic field}\label{dpB}

\begin{figure}[hb]
\epsscale{.7}
\plotone{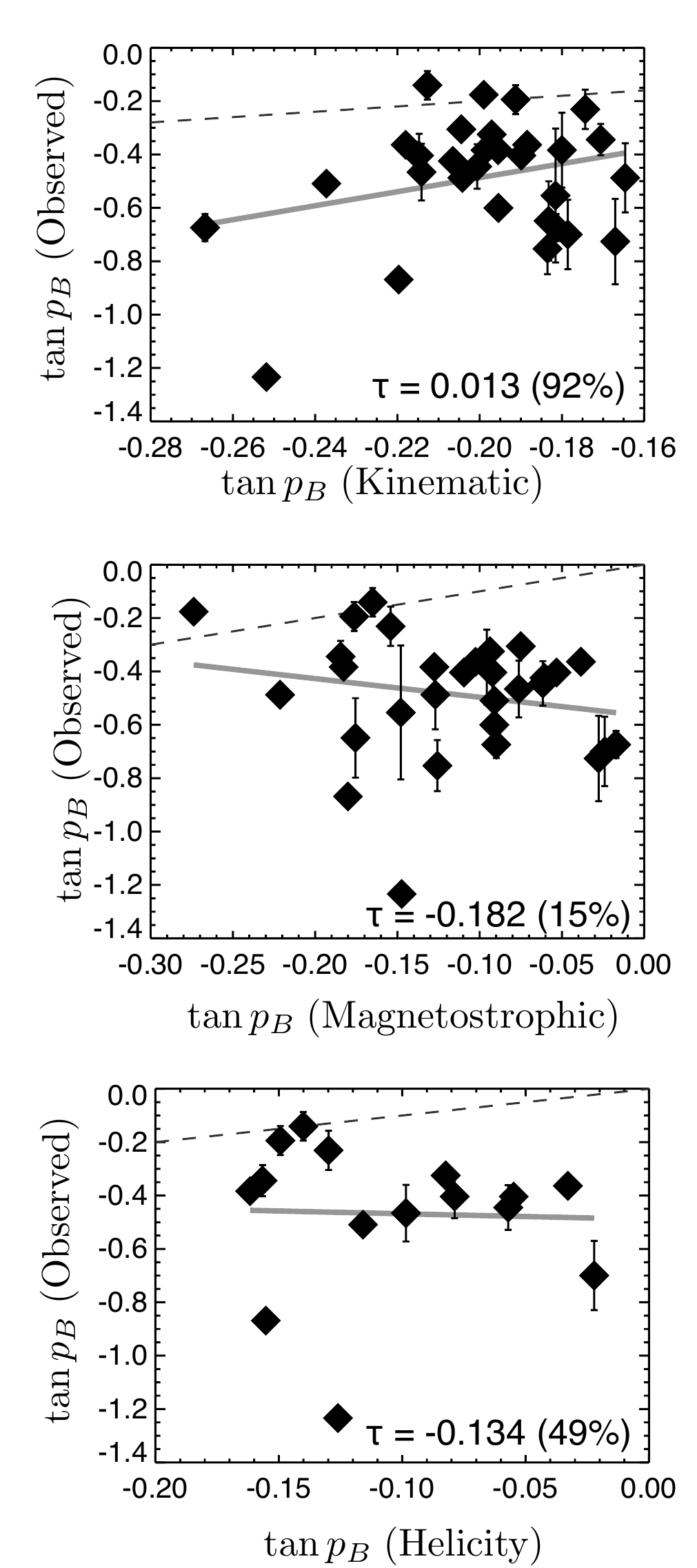}
\caption{\label{figuredynamopitch}
Comparison of the magnetic pitch angles observed and those predicted by:
the kinematic dynamo (top),
nonlinear dynamo with magnetostrophic saturation (middle),
nonlinear dynamo with saturation through helicity balance (bottom).
The gray lines are the best-fit linear relationship, while the black line
corresponds to the perfect agreement (unit slope).}
\end{figure}

To test theoretical predictions for the pitch angle of the mean magnetic field,
Eqs.~\eqref{pitch_kinematic_obs}, \eqref{pitch_strophic_obs} and \eqref{pitch_z}, 
we can use Pearson's linear correlation coefficient 
as we expect (or hope) to have a simple linear relation between the observed and
predicted values. The resulting correlation coefficients are shown in 
Table~\ref{tablestrengthcorrelations} and the scatter plots, in Figure~\ref{figuredynamopitch}.  
The predicted pitch angles for a kinematic dynamo and for the magnetostrophic balance 
depend on the rotation curve alone, and so can be tested for more galaxies
than the magnetic helicity balance model which includes more parameters.  
The kinematic dynamo, magnetostrophic balance 
and helicity balance model, all produce weak anti-correlations instead of
positive correlations.
All three models tend
to underestimate the magnitude of the observed pitch angles and 
both Eq.~\eqref{pitch_kinematic_obs} and Eq.~\eqref{pitch_z} predict
narrower ranges of $p_B$ than that observed.

%----------------------------------------------------------------------------------
% FIGURE 3:
%----------------------------------------------------------------------------------

The fact that Eq.~\eqref{pitch_kinematic_obs}, obtained from a kinematic dynamo 
model, does not agree with the observed magnetic pitch angles, 
can just mean that the mean-field dynamos in most, if not all, of the sample galaxies have
already entered a non-linear (saturated) stage. 
The failure of the magnetostrophic balance
model to predict the observed pitch angles suggests either that the force balance
involved is not the dominant mechanism of the mean-field dynamo saturation or
that most or all of the galaxies have a significant outflow across the disc-halo 
interface.
Either way, this 
implies that $p_B$ should depend on
the gas density and 
star formation rate: both $\Sigma$ and $\Sigma\SFRs$ appear in Eq.~\eqref{pitch_z}
but not in Eqs~\eqref{pitch_kinematic_obs} and \eqref{pitch_strophic_obs}.

The pitch angles predicted by the magnetic helicity balance model are also much 
smaller than those observed. This is because the 
actual values of $|R_{\omega}|$ in Eqs.~\eqref{pitch_strophic_obs} and \eqref{pitch_z}
exceed its normalization value: 
the strong shear leads to a prediction of a tightly wound up field with small pitch 
angles. So the dynamo models we have constructed overemphasize the role of shear in 
determining the mean magnetic field pitch angle.

%------------------------------------------------------------------------------------

%-------------------------------------------------------------------------
\section{The pair-wise correlations}\label{BCR}

\input{table5.tex}
%---------------------------------------
% Table 5 here 
%---------------------------------------

Apart from those for the few nearby galaxies discussed above, the data are very 
inhomogeneous, especially regarding the diversity of 
approaches to obtain magnetic field strength. Various authors use either
minimum-energy or pressure equipartition estimates to derive the total magnetic field
strength from synchrotron intensity. The strength of the mean magnetic field
is obtained from the polarized synchrotron intensity, most often without proper regard of 
depolarization. The estimates are often published without clear indication of the region
within the galaxy which they refer to. Moreover, the routinely used assumption of a direct local
relation between the cosmic-ray and magnetic energy densities or pressures is likely
to be wrong \citep{SSFBLT12}, leading to an overestimated $\BB$ if the linear resolution
of the observations is finer than a few hundred parsecs.

The consequences of the undesirable diversity in the existing interpretations of the
radio continuum observations of spiral galaxies, and way to improve the situation, become
evident as soon as one attempts to compare the data and develop a coherent picture
of magnetic fields in the galaxy sample. Therefore, we present here the pair-wise
correlations between magnetic and other parameters of the galaxies, being aware that
their physical significance may be limited.

We calculated Kendall's correlation coefficient $\tau$, given in Table~\ref{tabletaucoeffs},
for each pair-wise combination of magnetic field with other ISM parameters. Also shown 
in the table are
the significance levels and the number of galaxies for which we have the
required data.  
We chose to use Kendall's correlation coefficient, instead of the more
commonly used Pearson's product-moment correlation coefficient, 
as Kendall's coefficient is sensitive to nonlinear correlations, and we had
no \textit{a priori\/} reason to assume that any correlations are linear.  The
significance level $\nu$ represents the chance that a correlation
could be produced by uncorrelated data; lower significance
levels indicate that an observed correlation is more likely to be real.
We used the $\nu=5$\% to idenitfy statistically significant correlations.
This leads to ten significant results from the combinations tested.  

\begin{figure}[hb]
\epsscale{1.2}
\plotone{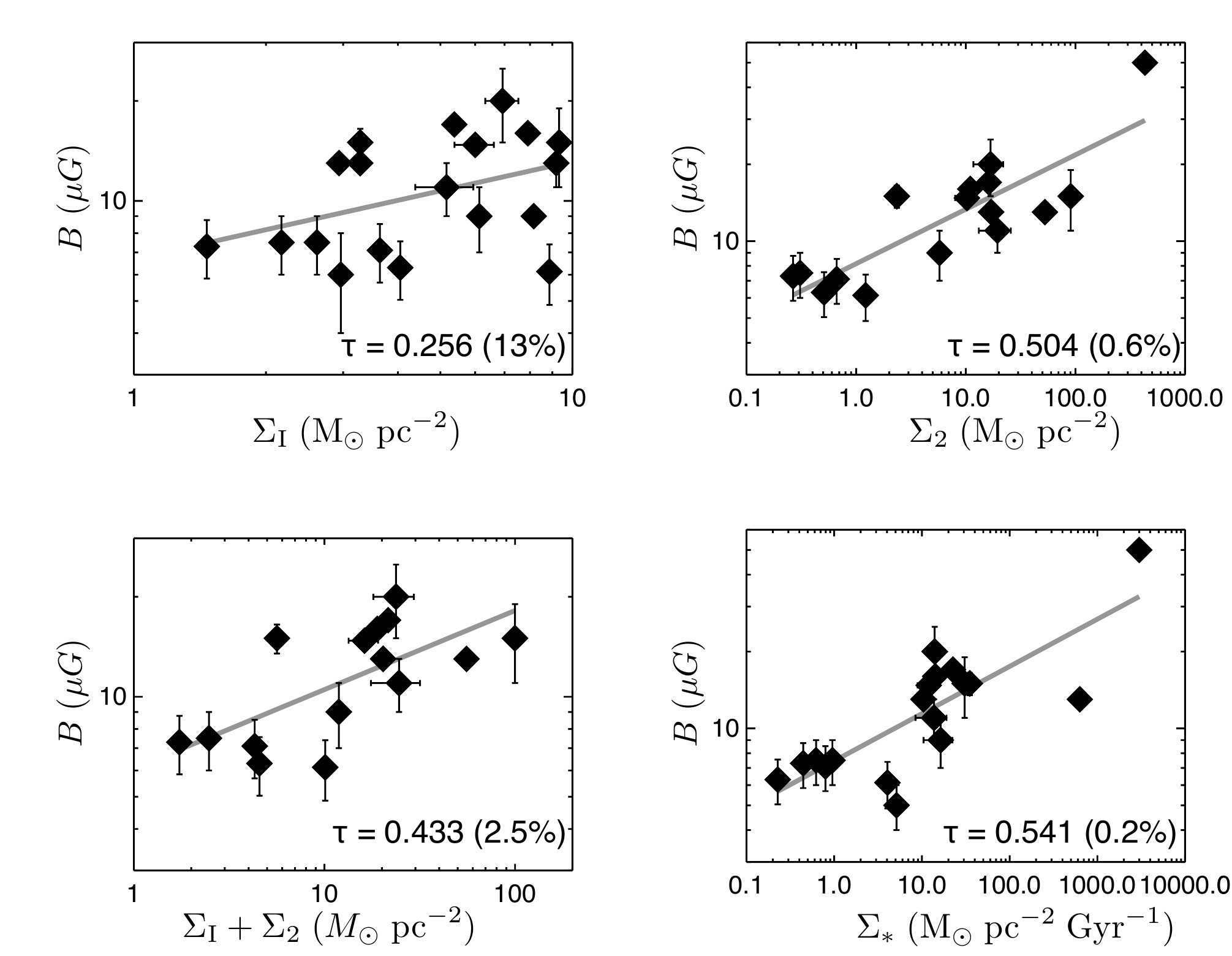}
\caption{\label{fig:firstcorrelations}
The total magnetic field strength $\B$ versus 
\ion{H}{I} surface density (top left),  
H$_2$ surface density (top right), 
total gas surface density (bottom left),
surface density of star formation rate (bottom right).  
The gray lines show the best-fit regression, with the error bars are included 
where they could be obtained from the original data. The correlation coefficient, 
$\tau$, and significance level $\nu$ (in brackets) are given in each panel.
}
\end{figure}

\begin{figure}
\epsscale{1.2}
\plotone{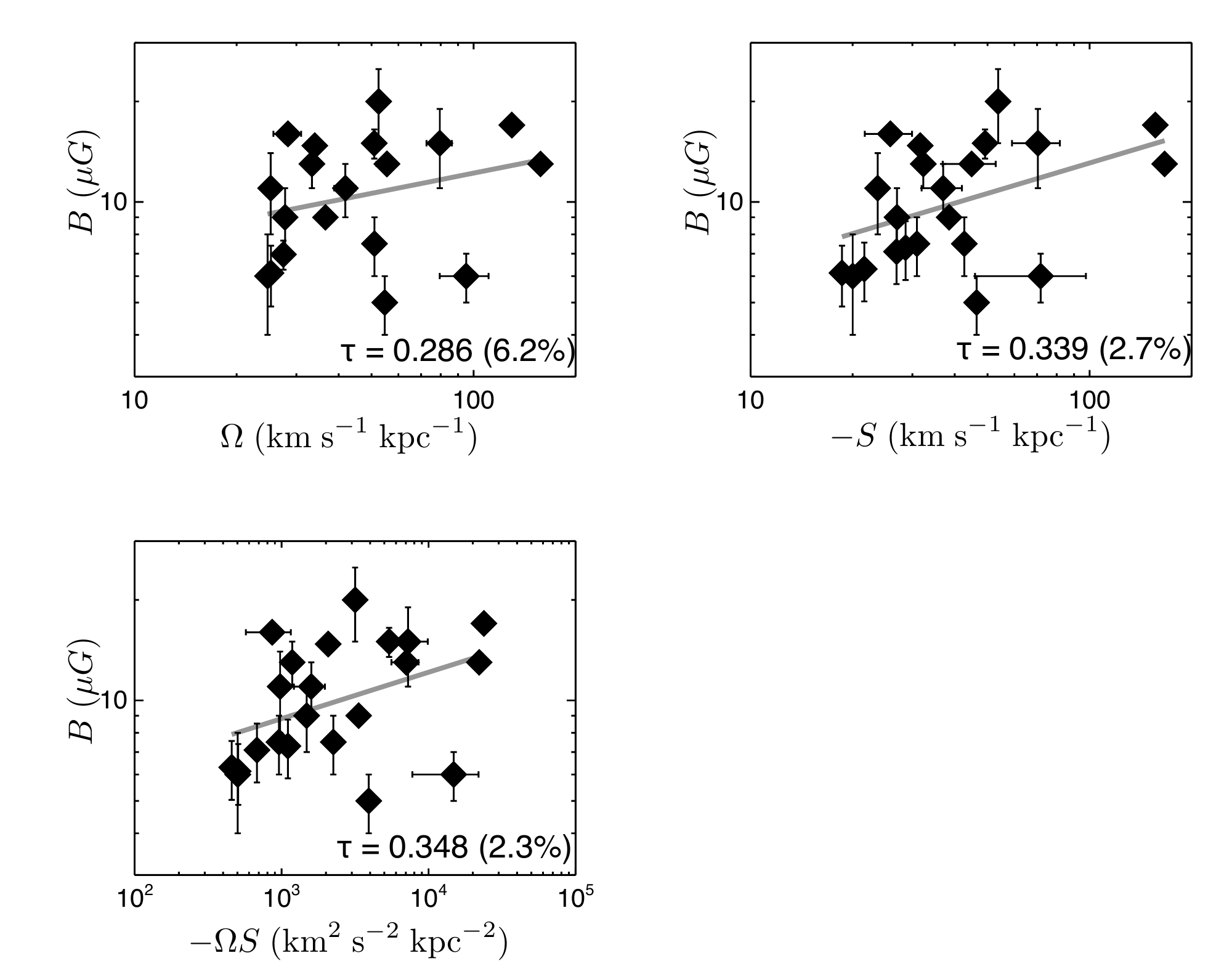}
\caption{\label{BvsRot}
As in Fig.~\ref{fig:firstcorrelations}, but for
the total magnetic field strength $\B$ versus
angular velocity (top left),
rotational shear rate (top right), 
and their product (bottom left).}
\end{figure}

\begin{figure}
\epsscale{1.2}
\plotone{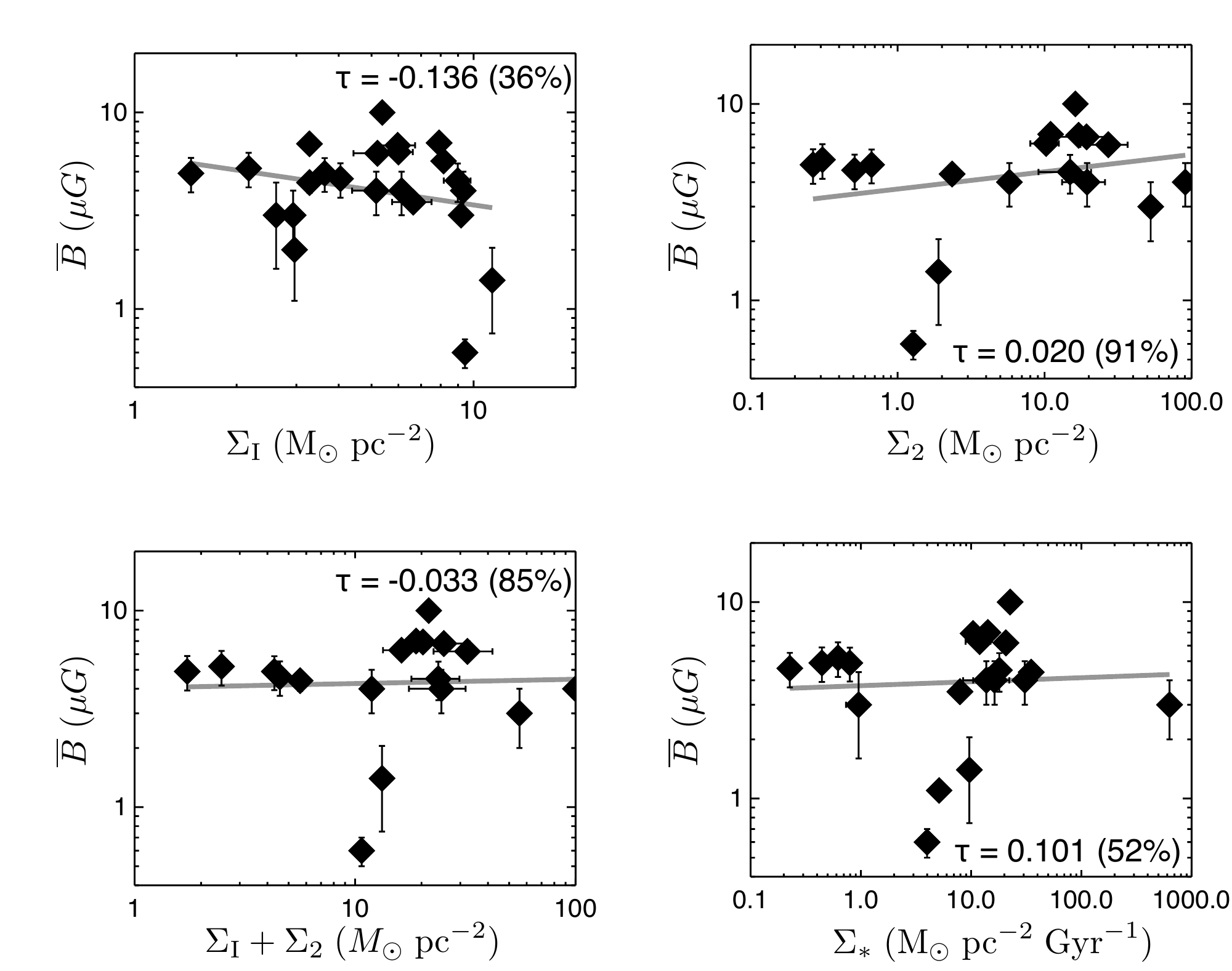}
\caption{\label{B0vsGas}
As in Fig.~\ref{fig:firstcorrelations}, but for the mean magnetic field strength
$\BB$.
}
\end{figure}

\begin{figure}
\epsscale{1.2}
\plotone{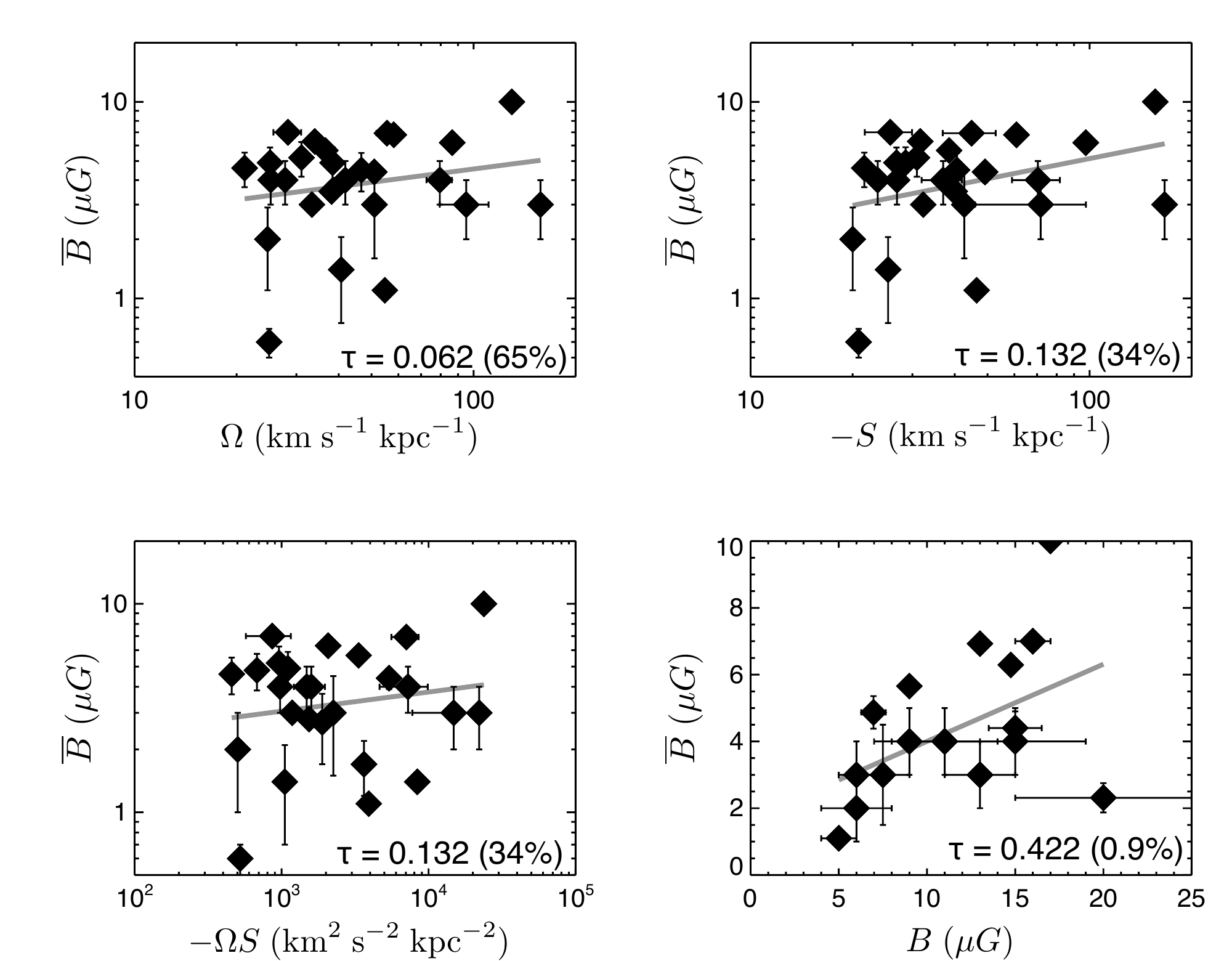}
\caption{\label{B0vsRot}
As in Fig.~\ref{BvsRot} but for the mean magnetic field strength
$\BB$, including its relation to the total magnetic field
strength $\B$ (bottom right).
}
\end{figure}

\begin{figure}
\epsscale{1.2}
\plotone{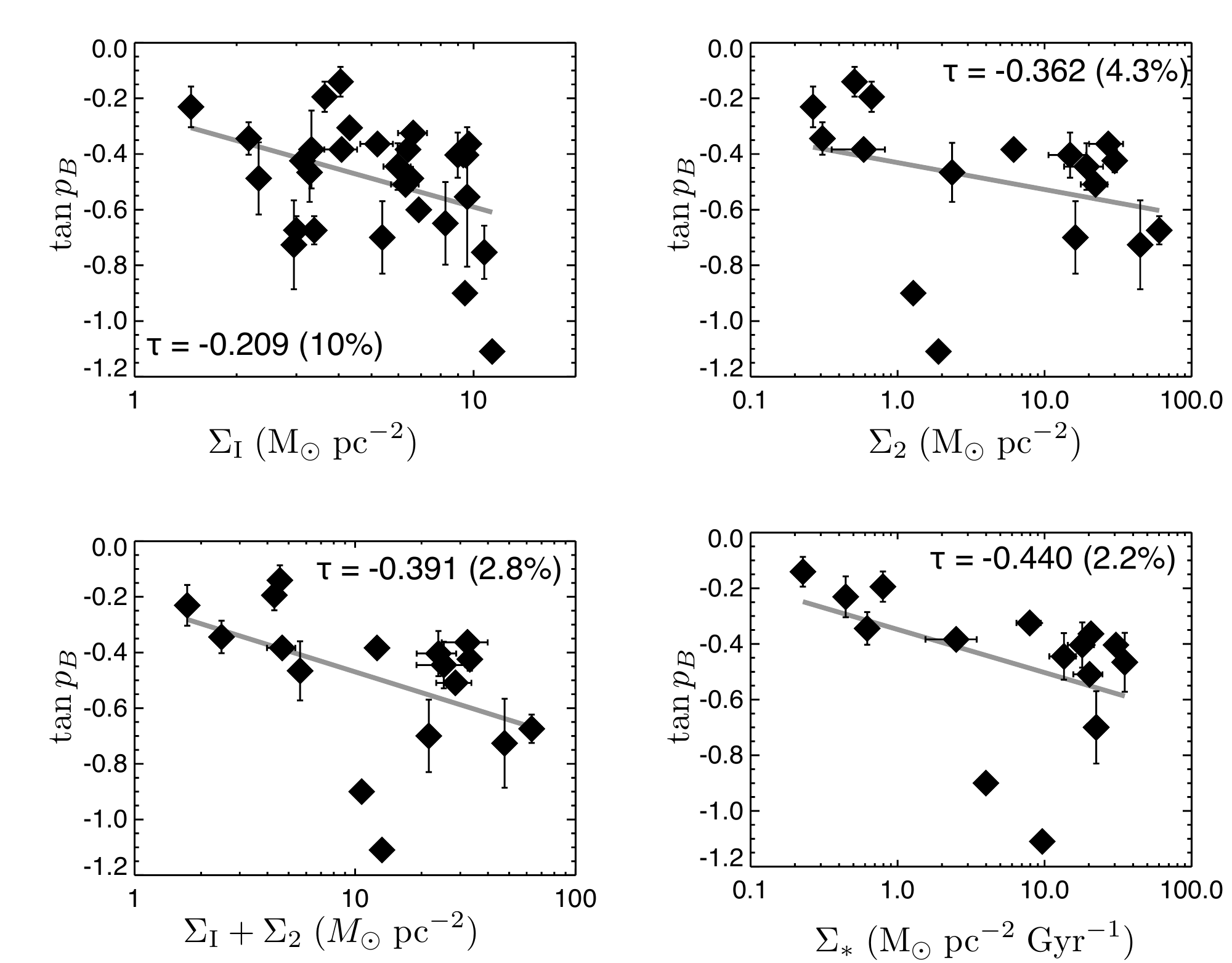}
\caption{\label{pBvrho}
The pitch angle of the mean magnetic field versus $\Sigma_\mathrm{I}$ (top left),
$\Sigma_2$ (top right), total gas surface density $\Sigma_\mathrm{I}+\Sigma_2$ (bottom left), and
$\Sigma\SFRs$ (bottom right). The gray lines show the best-fit regression, with the error bars are 
included where they could be obtained from the original data. The correlation coefficient, $\tau$, 
and significance level $\nu$ are given in each panel.
}
\end{figure}

\begin{figure}
\epsscale{1.2}
\plotone{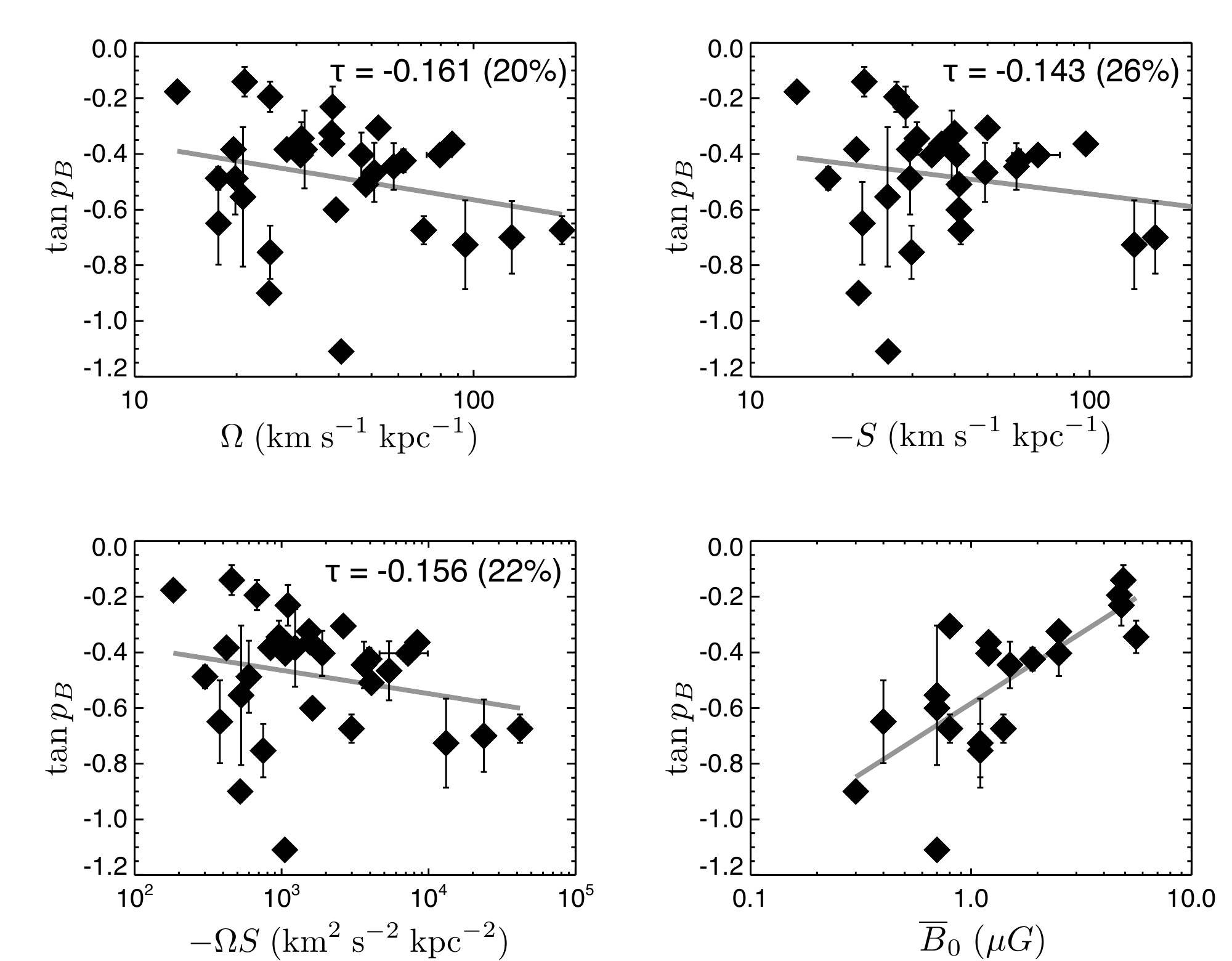}
\caption{\label{fig:lastcorrelations}
As in Fig.~\ref{pBvrho}, but for $p_B$ versus
angular velocity (top left),
rotational shear rate (top right),
their product (bottom left), and the axisymmetric component of the large-scale 
magnetic field (bottom right).}
\end{figure}

A summary of the pair-wise correlations is presented in Table~\ref{tabletaucoeffs} 
and the scatter plots and fits are shown in 
Figs~\ref{fig:firstcorrelations}--\ref{fig:lastcorrelations}.

%-----------------------------------------------------------------------
\subsection{Magnetic field strength}\label{pwMFS}

The total magnetic field strength $\B$ is significantly correlated with the
surface densities of H$_2$ and star formation rate. However, these variables 
are themselves correlated through the Schmidt--Kennicutt law, so the two 
correlations are not independent. The correlation with the star formation rate is 
similar to that found by \citet{Chyzy08} and \citet{Chyzy11}. The co-variation of $\B$ and
$\Sigma_2$ is consistent with Eq.~\eqref{rhon} with $k\approx0.21$, which
does not fit any of the standard models such as isotropic, two- or one-dimensional
compression. Since $k=0.21$ is far below the smallest of the values arising from those models, $k=1/2$,
the difference cannot be explained by a mixture of various behaviors but rather suggests
that either
scaling of magnetic field strength with the gas density alone,
Eq.~\eqref{rhon} fails to capture the relevant physics of the cloud formation, 
or that Eq.~\eqref{rhon} is not applicable to data averaged on scales of about one kiloparsec
(a typical linear resolution of the radio observations used).

The dependence expected from Eq.~\eqref{rhov} is recovered if the internal velocity 
dispersion depends on the cloud density approximately as $v\propto\rho^{-1/4}$. This 
is not inconsistent with Larson's laws \citep{L81}, first obtained for molecular clouds, but then
extended to diffuse clouds: $\rho\propto R^{-1.15\pm0.15}$ and
$v\propto R^{0.4\pm0.1}$ \citep{V-SB-PR97},
which can be combined to $v\propto\rho^{0.35\pm0.10}$. Together with Eq.~\eqref{rhov}, 
these relations imply $\B^2\propto\rho^{0.3\pm0.1}$, which is consistent
within $2\sigma$ ranges with $\B^2\propto\rho^{0.5\pm0.1}$ obtained here
despite the fact that individual clouds are not resolved in the data used.
 %%%%%%%%%%%%%%%
% Figure 4 - 7 
%%%%%%%%%%%%%%%

%----------------------------------------------------------------------
% FIGURE 4:
%----------------------------------------------------------------------

%----------------------------------------------------------------------
% FIGURE 5:
%----------------------------------------------------------------------

%----------------------------------------------------------------------
% FIGURE 6:
%----------------------------------------------------------------------

%----------------------------------------------------------------------
% FIGURE 7:
%----------------------------------------------------------------------

%---

The strength of the mean magnetic field $\BB$ exhibits no discernible pair-wise 
correlation with any of the simple variables tested.  This is somewhat surprising,
especially the lack of correlation with the rotational shear rate $S$ as one
can be quite confident that galactic differential rotation does affect the large-scale
magnetic field. It may be the case that the differential rotation, while necessary for the operation of the dynamo, is not the factor that limits the efficiency of the dynamo and ultimately is not dominant in determining the final magnetic field strength. If so, the correlation between rotation and the field strength may be too weak to be identified at 
reasonable significance, especially with the limited data sets currently available.

%--------------------------------------------------------------------
\subsection{Pitch angle of the mean magnetic field}\label{pwpB}

The pitch angle of the mean magnetic field is correlated, at better 
than 5\% significance, with the surface density of molecular hydrogen, 
the total gas surface density, and the star formation
rate. These three quantities are not independent, as the H$_2$ density is a 
substantial part of the total gas density, and the gas density and SFR are related by the
Schmidt--Kennicutt law. Of the three variables, $\Sigma\SFRs$ exhibits the
strongest connection with $p_B$, which could imply that the pitch angle is  
physically related to star formation, whereas the other, weaker, 
correlations reflect relationships between those parameters
and the SFR. To clarify this, one would require a more sophisticated statistical
analysis and, most importantly, more data.  

The tightest and perhaps the most remarkable close correlation that we find is that 
between the strength of the axisymmetric component of the mean magnetic field $\BB_0$ 
and the magnetic pitch angle. As $\BB_0$ gets stronger $p_{B}$ gets smaller: in other 
words, strong axisymmetric mean fields are more tightly wound than weak ones. 
Since $\BB_0$ has the fastest growth rate of the azimuthal magnetic field patterns 
according to galactic dynamo theory \citep{RSS88} this correlation may provide a way to 
rank the efficiency of the mean-field dynamos in different galaxies, with smaller $p_{B}$ 
indicating a more efficient dynamo. However, the interpretation of this correlation should 
be treated with caution until a predictive link between $\BB_0$ and $p_{B}$ is established. 
Our experience reported in Sect.~(\ref{TGDM}) suggests that this might be more difficult 
than one might naively expect.

Perhaps surprisingly, $p_B$ shows no significant correlation with the rotational shear
$S$ despite the fact that a mean magnetic field must be affected 
by the galactic differential rotation. We note, however, that the weak negative 
correlation of $|p_B|$ with $S$ ($\Omega$ and $S=r\,\dd\Omega/\dd r$ are 
functionally related) suggests a decrease in $|p_B|$ as $S$ increases,
consistent with a tighter winding of magnetic spirals under stronger 
differential rotation.

\begin{figure}
%\epsscale{0.7}
\plotone{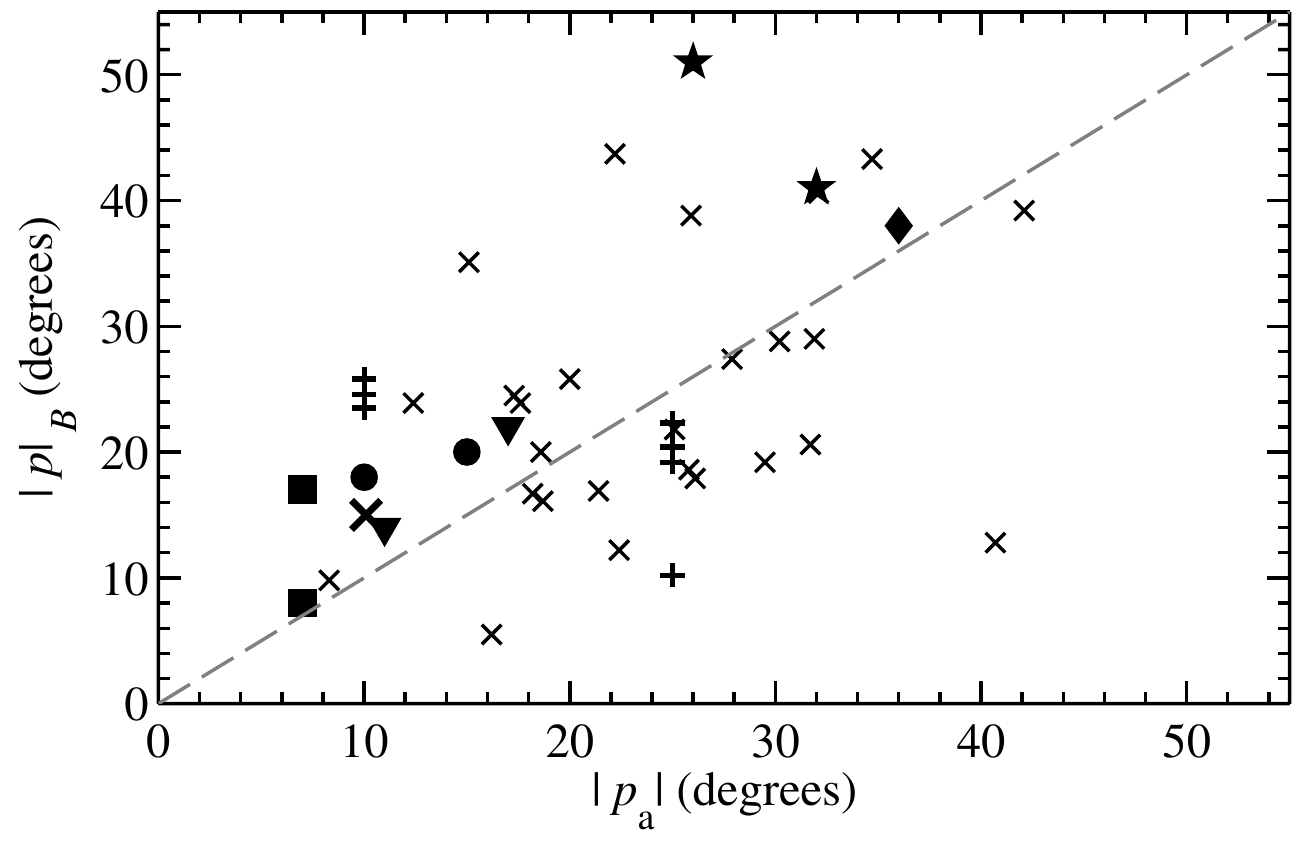}
\caption{\label{pa_pB}
As in Fig.~\ref{pBvrho}, but for $p_B$ versus the pitch angle of spiral arms
$p_\mathrm{a}$ in a selection of galaxies with the data available. 
M51 (crosses): the local pitch angles of the mean magnetic field and the 
CO spiral arms \citep[from Fig.~9 in][]{PFSBBFH06}.
M33 (stars): $p_\mathrm{a}$ from \citet[Table~1 of][]{SH80} (measured from the local 
circumference rather than the local radius as in the original) and $p_B$ from the 
$m=0+2$ fits of \citet[][, Table~2]{Tabatabaei08}, both averaged in the radial ranges 
1--3\,kpc and 3--5\,kpc. 
IC~342 (plus signs): $p_B$ from Beck (2014, private communication, obtained by
averaging the pitch angles of the $B$-vectors at $\lambda6\cm$ in sectors of 
$10^\circ$ wide in azimuth and $2\kpc$ wide in radius in the galaxy plane)
and $p_\mathrm{a}$ from \ion{H}{I} observations of \citet{CTH00}.
M31 (squares): $p_B$ from \citet{Fletcher04}, $p_\mathrm{a}$ from \citet{Nieten:2006}.
M51 (circles): $p_B$ from \citet{Fletcher11}, $p_\mathrm{a}$ from \citet{BHKNPSS97}.
M81 (triangles): $p_B$ from \citet{Krause89}, $p_\mathrm{a}$ from \citet{Oort:1974}.
NGC~6946 (diamond): $p_B$ and $p_\mathrm{a}$ from \citet{Frick:2000}.
The Orion arm of the Milky Way (large cross): $p_B$ from \citet{FSSS01},
$p_\mathrm{a}$ from \citet{XLRMZBMDZ13}.
The straight line corresponds to $p_B=p_\mathrm{a}$.}
\end{figure}

As with the strength of the mean magnetic field, a 
possible
explanation
of the weak correlation of $p_B$ with $S$ is that the pitch angle depends not
only on $S$ but also on other parameters such as the SFR, as revealed here. 
In Section~\ref{TGDM} we investigated the correlation of $p_B$ with different 
combinations of parameters derived from specific physical models in Section~\ref{ACDI},
but this does not improve the correlation. It is more plausible that the
magnetic pitch angle is affected by the spiral arms as discussed in Section~\ref{CSMF}.
Figure~\ref{pa_pB} demonstrates a clear correlation between the pitch angles
of the spiral arms, $p_\mathrm{a}$, and the mean magnetic field, $p_B$.
Grey crosses show the local pitch angles in M51 obtained  by \citet{PFSBBFH06}
(their Figure~9); these are the local pitch angles of the total magnetic field and the
CO spiral arm segments. The mean value of the difference of the pitch angles is close
to zero with the standard deviation of $10^\circ$ and the median value of $1.5^\circ$. 
The other symbols show
the pitch angles of the large-scale magnetic field and spiral arms in several 
galaxies specified in the figure caption. The mean and median values of the
difference between the pitch angles $|p_B|-|p_\mathrm{a}|$
(all data points in Figure~\ref{pa_pB} except for the local 
pitch angles in M51) are about $5^\circ$ (with the standard deviation of
$9^\circ$ and only four negative values of $|p_B|-|p_\mathrm{a}|$ out of 17,
all from the outer rings in IC~342). 
The global estimates of $p_B$ are plausible to
be biased to the interior of spiral arms because of the smaller errors of
the polarization angles in the arms where the polarization intensity is 
generally stronger.

%%%%%%%%%%%%%%%
% FIGURE 8 & 9 
%%%%%%%%%%%%%%%

%----------------------------------------------------------------------
% FIGURE 8:
%----------------------------------------------------------------------

%----------------------------------------------------------------------
% FIGURE 9:
%----------------------------------------------------------------------

The pitch angles of spiral arms are most often obtained
from a global fitting of a logarithmic spiral whereas magnetic pitch angles are 
obtained in a less restrictive manner and are allowed to vary 
with radius (and often, with azimuth). This is likely to affect the relation 
between the pitch angles discussed above. The local magnetic pitch angles in M51 
are, on average, very close to those of the CO spiral arms, but the former are
derived from the polarization angles that are known to be affected by anisotropic
random fields \citep{SSFBLT12}; hence, the comparison should be treated cautiously as it may
not reflect reliably the relation between the local pitch angles of the 
\textit{large-scale} magnetic field and spiral arms. Despite these caveats,
we can conclude that there is a tight correlation between the pitch angles 
of magnetic field and spiral arms. Nevertheless, they differ systematically, 
with the integral lines of the large-scale magnetic field likely to be more 
open than the spiral arms, $|p_B|-|p_\mathrm{a}|\simeq5\text{--}10^\circ$. 
 
 %%%%%%%%%%%
% FIGURE 10
%%%%%%%%%%%
 
 %---------------------------------------------------------------
%FIGURE 10: 
%---------------------------------------------------------------

%----------------------------------------------------------------------
\subsection{Comparison with earlier results}\label{CER}
The local relationship between the total magnetic field strength and the
star formation rate in the galaxy NGC~4254 was studied by \citet{Chyzy08}
who used 277 locations in the galaxy, all separated by 1.2\,kpc. He found a
strong correlation between the total magnetic field strength and the star
formation rate, 
$B\propto\Sigma_*^k$, with Pearson's correlation coefficient of $r=0.93$ and 
 $k=0.18\pm0.01$.
Our analysis of the global averages for these quantities over many spiral
galaxies gave Kendall's correlation coefficient of 0.49 and a significance
level of 1.1\% with 15 galaxies, confirming strong correlation 
(see the bottom-right panel of Figure~\ref{fig:firstcorrelations}).  
The close agreement between our results and those of \citet{Chyzy08} 
is more easily seen when we calculate 
Pearson's correlation coefficient for our data; we obtain
$r=0.77$ (at significance level 0.09\%) for the logarithmic 
variables, with 
$k=0.20\pm0.05$. Like us, \citet{Chyzy08} 
found no correlation between the mean magnetic field and the star formation rate.

\citet{TBFBS13} find a similar correlation in NGC~6946 with $k=0.14\pm0.01$. These
authors attribute the difference from \citet{Chyzy08} to the
special conditions in NGC~4256 that belongs to the Virgo cluster.
However, our results suggest that NGC~6946 may be anomalous in this respect.

\citet{Chyzy08} also analyzed the field regularity $\BB/\Bb$ in NGC~4254 as a 
function of the local star formation rate, to find an anticorrelation
$\BB/\Bb\propto\Sigma\SFRs^{-0.32\pm0.01}$ with $r=-0.71$. 
For the sake of comparison, we also considered $\BB/b$ to obtain
$r=-0.44$ (at 13\% significance level) using logarithmic variables. 
The value of Kendall's correlation coefficient is
$\tau=-0.21$ (at 32\% significance, 13 galaxies). The best-fit
power law has the form $\BB/\Bb\propto \Sigma\SFRs^{-0.15\pm0.09}$, 
significantly shallower than that in NGC~4254.

\citet{TBFBS13} also find a rather weak anticorrelation of $\BB/\Bb$ and
$\Sigma_*$ in NGC~6946, with the dependence being nearly flat at 
$\Sigma_*\lesssim0.01\,M_\odot\yr^{-1}\kpc^{-2}$ but somewhat steeper at larger
star formation rates.

It is possible that the difference between these results arises because the 
relation of $\BB/\Bb$ to $\Sigma\SFRs$ varies widely between galaxies
in our sample. It is more plausible, however, that it might be caused by the fact 
that we use the values of $\BB$ and $\Bb$ averaged over relatively large regions 
in galactic disks (several square kiloparsecs), whereas Chy{\.z}y used values 
averaged over the beam area (about $1.5\kpc^2$). Then the difference 
 is understandable if $\BB/\Bb$ and/or 
$\Sigma\SFRs$ vary significantly at
a scale of order a few kiloparsecs. Indeed, from Figure~7b of \citet{Chyzy08},  
$0.2\lsim \BB/\Bb\lsim 1.6$ and $\Sigma\SFRs$ varies by a factor of 300 between 
various locations across NGC~4254.

\citet{Chyzy11} compared magnetic field strengths to the global averages
of various galactic parameters for a sample of seven dwarf galaxies.  
They found a correlation $\B\propto \Sigma\SFRs^{0.30\pm0.04}$ with
the correlation coefficient of 0.94. 
\citet{HBLHBBB14} found a similar relation for
a sample of 17 galaxies (containing two dwarf galaxies) assuming energy
equipartition between cosmic rays and magnetic fields at 1\,kpc scale. 
This variation is steeper than that obtained here, $B\propto \Sigma\SFRs^{0.19\pm0.03}$,
and by \citet{Chyzy08}.

\citet{Chyzy11} also found a correlation coefficient of 0.78 between the total
magnetic field and the \ion{H}{I} surface density in their sample of dwarf galaxies, with 
a power law  exponent of $0.47\pm0.09$.  Our selection of spiral galaxies 
showed no correlation of $\B$ with the \ion{H}{I} density, but a significant correlation 
with the density of the molecular gas. 
\citet{TBFBS13} also note that the synchrotron intensity in NGC~6946 is only
weakly correlated with \ion{H}{I} density, but find a significant correlation with
the total gas density, $B\propto(\Sigma_1+\Sigma_2)^m$ with $m=0.23\pm0.01$, evidently
dominated by a correlation with molecular gas, $\Sigma_2$. Our results are in a good
agreement, with $m=0.24\pm0.07$. 

To conclude, there is a reasonable degree of agreement between the results 
presented here and earlier work; where there is disagreement this can be plausibly 
understood as arising from differences in the number and/or type of galaxies in the 
samples or from differences in the nature of the data used. This gives us some
confidence in the results
presented here despite the 
limited size and quality of the data available.

%---------------------------------------------------------------------------
\section{Discussion}\label{Disc}

Earlier attempts to establish correlations between galactic magnetic fields
and other galactic parameters met with quite limited success. Among a few
notable exceptions is the correlation of magnetic field strength in dense
interstellar clouds with gas density \citep[e.g.,][]{Troland86}
or, rather, kinetic energy density \citep{B00}. There are claims
of a correlation of the orientation of magnetic fields in molecular clouds
with that of the magnetic field in the surrounding kiloparsec-wide region \citep{Han07,Li11}.
If confirmed, such a correlation would be difficult to explain since  
random magnetic fields at a scale of order $100\p$, comparable to
the size of a region which collapses to form a molecular cloud, are several times
stronger than the large-scale magnetic field \citep[see ][for a
review]{SSFBLT12}. Thus, any correlation of magnetic field orientations within a 
cloud and at a scale of order kiloparsec would require a special explanation
which is not immediately obvious.

The general properties of large-scale galactic magnetic fields
strongly suggest that they are formed by mean-field dynamo action
\citep{S07}. However, detailed, quantitative comparisons with the dynamo 
theory still remain mostly restricted to individual galaxies 
\citep[especially barred galaxies: ][]{Beck05,MSESBS07}. 
Here we have made the first steps towards such a comparison in a sample of 
20 spiral galaxies. 
There are numerous caveats to our analysis:
\begin{itemize}
\item[--] The sample of galaxies used is far from being statistically 
significant, even though it contains all of the galaxies for which the required data have 
been published in sufficient detail.
 
\item[--] The sample is very inhomogeneous, with different morphological classes of galaxy
and widely varying linear resolution of the radio 
observations used to estimate their magnetic properties.

\item[--] The observational magnetic field estimates themselves are derived using two different 
methods (either from energy or pressure equipartition between magnetic fields and cosmic rays 
or from the Faraday rotation) 
that involve a range of further assumptions about the ISM properties in each galaxy,
often involving presumed (but not confidently established) similarity with the Milky Way. 
In particular, estimates of magnetic field
strength derived from the total and polarized radio emission, when applied at scales
less than about a kiloparsec, lead to a systematically underestimated
random magnetic field and, correspondingly, an overestimated mean magnetic
field \citep{SSFBLT12}. Correlations (or anti-correlations) in the spatial distributions of 
the relativistic and thermal electrons and the interstellar magnetic 
fields can also significantly affect the magnetic field estimates, which are generally
derived assuming 
that magnetic fluctuations are statistically independent of those
in the number density of thermal electrons \citep{BSSW03}.
Anisotropy of the random magnetic field produced by the rotational shear and
inhomogeneous outflows can affect estimates of the mean magnetic
field strength from the degree of polarization \citep{SBSBBP98,Beck05} and 
cause discrepancies between estimates obtained from polarized intensity and
Faraday rotation \citep{Fletcher11, SSFBLT12}.

\item[--] When testing the theoretical predictions for the strength and pitch angle of the large-scale
magnetic field, we neglected the variation of the disk thickness with galactocentric radius.
However, disk flaring can have a strong affect on the magnetic pitch angle. The
correlations discussed in Section~\ref{dpB} may become stronger if
this factor could be taken into account.
\end{itemize} 

These complications hamper any statistically rigorous
testing of the theoretical results \textit{with the observational data available.}
Some of the problems could be resolved in the near future by a careful re-reduction of the
existing observations and a systematic use of the existing theoretical results
\citep[e.g., ][]{Beck05}, the others require additional observations and theoretical work.

%----------------------------------------------------------------------------------
\section{Summary and Conclusions}\label{SaC}

We have explored correlations between magnetic fields (both random and mean) 
and the gas density (both atomic and molecular), 
star formation rate, angular velocity and rotational shear in a sample of 20 
spiral galaxies. Both pair-wise
correlations between the variables and the dependence of the mean magnetic field
on parameter complexes suggested by various dynamo models have been tested.
The size and statistical quality of the sample are limited, but several
correlations are significant and admit physical interpretation.

We found that the \textit{total\/} magnetic field strength (dominated by the 
random magnetic field in most cases) is significantly correlated with
the molecular gas density and the star formation rate, and has a weak or no
pair-wise correlations with the atomic gas density and rotational
parameters. The correlation with gas density and the lack of correlation with
the rotational parameters is not surprising. The relation between
gas density and magnetic field strength is shown to be 
marginally
consistent with 
Larson's relations between the size, density
and velocity dispersion of interstellar gas clouds.

The mean magnetic field strength $\BB$ does not exhibit any significant
pair-wise correlation with any of the individual galactic parameters tested, 
even the rotational shear rate $S$. Correlations of $\BB$ and its pitch angle $p_B$ with molecular gas 
density and star formation are only modest, but still stronger than with 
$S$. This may appear surprising as there is no 
obvious direct physical connection between $p_B$ and star formation rate.
We interpret this as an indication that the physical connection
is between $\BB$ or $p_B$ and combinations of the galactic parameters: we
turn to the relations predicted by galactic mean-field dynamo theory (other
theoretical ideas on the origin and evolution of galactic magnetic fields do not
offer testable predictions).

We have found an unexpected close correlation between the strength of the axisymmetric 
component of the mean magnetic field and the magnetic pitch angle. This would be 
easily explained if we had found a tight correlation between the shear $S$ and $p_{B}$ 
--- strong shear produces efficient mean-field dynamos and preferentially generates 
azimuthal field over radial field, resulting in a small pitch angle ---  but we did not. 
If a well-founded theoretical explanation for this correlation can be uncovered then it 
may provide an extremely useful diagnostic for the mean fields of galaxies, as the 
pitch angle is far easier to determine than the strength of the mean magnetic 
field: it can be directly measured whereas field strength can only be indirectly inferred.

The recent developments in nonlinear galactic dynamos, where
the steady state of the large-scale magnetic field is controlled by
magnetic helicity balance, indeed predict that
both $\BB$ and $p_B$ should depend on gas density and the intensity
of an outflow from the galactic disc (which can be either a fountain or winds) and, 
hence, on the star formation rate.

On the other hand, the pitch angle of the mean magnetic field is correlated with
the pitch angle of the spiral arms. We stress that
the two pitch angles still differ systematically by about $5^\circ$. 
Such a correlation arises naturally from a one-dimensional compression of
magnetic field in the spiral arms.

We have tested a range of predictions of galactic dynamo theory, from the
most general ones that only rely on their elementary features, to detailed
models based on specific mechanisms for dynamo saturation. The saturation models 
considered are: (i)~energy equipartition between the large-scale magnetic field
and the turbulent kinetic energy, (ii)~balance between the
Coriolis and Lorentz forces (the magnetostrophic balance), and (iii)~magnetic
helicity balance. 
In the galaxy sample used, the mean magnetic field strength
does not exhibit statistically significant correlations with the
parameter combinations corresponding to either of these models.
There can be many reasons for the lack of agreement, coming from flaws in both
theory and the data inferred from observations. Regarding the theory, not all
physical effects, even those relatively well explored, have been included into
the models explored. Such effects include radial flows which can affect the magnetic pitch angle
\citep{MSS00}, additional helicity fluxes which can enhance the mean magnetic
field in the helicity balance model \citep{VC01,VS14},
deviations from axial symmetry in the dynamo solutions, etc. Perhaps more importantly,
the outflow model of Appendix~\ref{OS} may be wrong or oversimplified; more work is required here. 
In favour of the models we used, they only contain more or less directly observable parameters
(unlike the more involved alternatives).

The data used here have been obtained from observations performed with different telescopes,
at different frequencies, resolutions and sensitivities. Furthermore, the only unifying feature
of the galaxies in the sample is that they are all disk systems; otherwise, they are
widely diverse. The problem with the observations can be alleviated with new 
surveys of representative galaxy samples or by 
reducing the existing observations to a common set of resolutions, sensitivities, etc.

The import of this paper is perhaps more the methods used
rather than the results. Particularly important aspects of our experience are: (i)~The need for
a physically motivated, simple and yet realistic nonlinear model of galactic magnetic fields.
Such a model ought to be calibrated using the best observational data in the Milky Way
and the nearest galaxies, and then carefully extended to a larger sample of galaxies.
(ii)~The need for a comprehensive, unbiased, statistically representative database
of galactic magnetic fields and of a broad range of relevant galactic properties. The current
surge of activity in galactic magnetism connected with the LOFAR and SKA projects
offers excellent opportunities in this area.

%----------------------------------------------------------------------------------
\acknowledgments
We are grateful to Rainer Beck for useful comments.
and for providing his results on IC~342 prior to publication. 
Detailed and useful comments
of the anonymous referee are gratefully acknowledged.
C.~V.~E.~would like to acknowledge the financial support provided by
\textit{Alberta Innovates}. J.~C.~B. acknowledges support from the Natural
Sciences and Engineering Research Council of Canada. A.~F.\ and A.~S.\ 
have been supported by the Leverhulme Trust (grant RPG-097) and the STFC (grant ST/L005549/1). 
We also acknowledge the
extensive use of NASA's Astrophysics Data System, and the SIMBAD Astronomical
Database operated by the Centre de Donn\'ees astronomiques de Strasbourg.
%----------------------------------------------------------------------------------
%----------------------------------------------------------------------------------

\newpage

\clearpage

%----------------------------------------------------------------
\appendix
\section{Correction for unresolved gradients in the mean magnetic field}
\label{UGRMF}
Our dataset is very diverse covering a wide range of galaxy types and
observational parameters. In some galaxies, the resolution was high enough to
resolve a significant part of the mean magnetic field, in others,
unresolved gradients could have significantly reduced the intensity of
polarized emission. In the latter case, a correction for unresolved
gradients, relying on a simple model of the large-scale magnetic field 
$\BBv$, can be introduced as follows. In the galaxy sample considered 
here, this correction was found to be insignificant.

We neglect the vertical component of $\BBv$ in comparison with the
horizontal ones, $|\BB_z|\ll |\BB_r|,|\BB_\phi|$ in the cylindrical 
reference frame $(r,\phi,z)$ with the origin at the galactic center
and the $z$-axis aligned with the rotation axis. The field strength 
is assumed to depend on the galactocentric radius and azimuthal angle $\phi$ as 
\[
\BB=B_0\exp(-r/R_B)\cos[m(\phi-\beta_m)]\,,
\]
where $m$ is the azimuthal wave number ($m=0$ corresponds to an axisymmetric
field, $m=1$ to a bisymmetric structure, etc.), $\beta_m$ is the phase of the
$m$'th mode, $R_B$ is the radial length scale of the field strength, and $B_0$
is the characteristic field strength. As shown by \citet{BHKNPSS97} (their Appendix~A), 
the Cartesian components $\BB_x$ and $\BB_y$ of the projection of a horizontal 
galactic magnetic field onto in the sky plane (with the $x$-axis aligned with 
the galaxy's major axis) are given by
\begin{equation}\label{BxBy}
\BB_x= \BB_r\cos\phi - \BB_\phi\sin\phi\,,
\qquad
\BB_y= (\BB_r\sin\phi+\BB_\phi\cos\phi)\cos i\,,
\end{equation}
where $i$ the inclination angle of the galactic disk ($i=0$ corresponds to
the face-on view). Using the pitch angle of the magnetic field $p_B$, we have
\[
\BB_r=\BB \sin p_B\;,
\qquad 
\BB_\phi=\BB \cos p_B\;.
\]
Tedious but straightforward algebra then leads to
\[
\BB_\perp^2=\BB_x^2+\BB_y^2
=\tfrac12 B_0^2\exp(-2r/R_B)\left[1+\cos^2 i -\sin^2i\,\cos 2(\phi-p_B)\right]
                        \cos^2[m (\phi-\beta_m)]\,.
\]
For the sake of simplicity, we only consider the case $m=0$ below.

When observed at a finite resolution, all the variables are
averaged across the beam, i.e., over the corresponding ranges of $\phi$ and $r$ in the
galaxy plane. The beam shape in the galaxy plane depends on
$\phi$ and $i$. If the angular resolution of the observations is $\theta$ and
the distance to the galaxy is $D$, the beam width at the minor axis is $2d/r$
along the azimuth $\phi$ and $2d/\cos i$ along $r$, where $d=D\tan i$.
Assuming that $d\ll r,R_B$ and keeping only the leading terms in $d/(r\cos i)$
and $d/R_B$, we obtain near the minor axis ($\phi\approx\pi/2$):
\begin{align}\label{BperpD}
\left\langle{\BB}^2_\perp\right\rangle&=(4D\cos i)^{-1}\int_{r-D/\cos i}^{r+D/\cos i}r'\,dr'
       \int_{\phi-D/r}^{\phi+D/r} \BB_\perp^2(r',\phi')\,\dd\phi'\nonumber\\
  &\approx\tfrac12 B_0^2e^{-2r/R_B} [1+\cos^2i-\sin^2i\, \cos2(\phi-p_B)]\,,
\end{align}
where $\left\langle{\BB}^2_\perp\right\rangle$ is ${\BB}_\perp^2$ averaged over the beam
area.

Near the major axis, $\phi\approx0$, the beam size is $2D$ along $r$ and
$2D/\cos i$ along $\phi$, and Eq.~(\ref{BperpD}) still applies for a narrow beam.

Remarkably, the result is independent of the beam size as
long as $d\ll r, R_B$ (such a dependence is in the higher-order terms in
$d/r$), and the main factor which affects the observed field magnitude via
unresolved gradients is the galaxy's inclination and azimuthal position
within the disk.
We have verified that the dependence on $m$ is also insignificant and, for our
purposes, we can use $m=0$ without any significant loss of accuracy.
Assumptions such as the cosmic ray--magnetic field equipartition or the statistical
independence of thermal and relativistic electron densities and magnetic field,
routinely used in deriving the field strength, introduce by far stronger uncertainties.

%----------------------------------------------------------------------------
\section{The outflow speed}\label{OS}
Both galactic fountains and winds are driven by supernova explosions that produce
hot gas. The speed of the hot gas at the base of the outflow averaged over the
disk surface, $V_z$ depends on the number of stars born per unit time per unit area, 
roughly estimated as $\Sigma\SFRs/M_*$, where $M_*$ is the average stellar mass. The 
supernova rate in a galaxy of a radius $R$ is given by 
\begin{equation}\label{SNrate}
\nu\SN=\delta\SN\pi R^2 \Sigma\SFRs/M_*\,,
\end{equation} 
where $\delta\SN\simeq8\times10^{-3}$ is the fraction of stars that evolve to 
supernovae (i.e., those in the mass range $10<M/M_\odot<40$) for the initial 
mass function of \citet{K08},
which also has $M_*=0.85 \,M_\odot$.

We present two estimates of the galactic outflow speed, one based on energy
conservation and the other, on a model of the break-out of a superbubble (produced by
an OB association) through the galactic layer of neutral gas. The dynamo action constrained
by magnetic helicity conservation only requires an outflow through the disc surface,
and it is unimportant whether it is a wind or a fountain. In the latter case,
the gas returns to the disc in the form of dense clouds. As the formation of the clouds
would involve intense reconnections, it is likely that the magnetically nontrivial structures 
are deposited in the halo rather than returned to the dynamo active disc. Thus, our estimates
of the outflow speed are independent of the depth of the galactic gravitational potential.
The superbubble break-out model of \citet{MLMC88}, used below, includes the
acceleration due to gravity.

%----------------------------------------------------------
\subsection{Energy conservation}\label{OS_EC}
The rate of supernova energy supplied per unit area follows as
$\delta\SN E\SN \Sigma\SFRs/M_\odot$, where $E\SN\simeq10^{51}\erg$ is the
supernova energy. If a fraction $\eta$ of the supernova energy feeds the outflow,
and the fraction of supernovae that occur in OB associations (and thus drive the 
outflow) is $\epsilon\SN$, the surface density of the energy supply rate to the 
outflow is
\[
\dot{E}\simeq \epsilon\SN \eta \delta\SN E\SN\Sigma\SFRs/M_*\,,
\]
and $\epsilon\SN\simeq0.7$ \citep{KH88}. This energy is carried away from the warm 
gas layer through both faces of the galactic disc at a time scale $h/V_z$, so that 
the energy in the disk is lost at a rate
\[
\dot{E}\simeq 2\rho\h V_z^3\,.
\]
The balance of the energy supply and loss rates in the disc leads to 
\begin{align}\label{VzE}
V_z\simeq&\left(\frac{\epsilon\SN\,\eta\,\delta\SN E\SN\Sigma\SFRs}{2\rho\h M_*}\right)^{1/3}\nonumber\\
=&90\,\frac{\!\km}{\!\s}
		\left(\frac{\epsilon\SN\,\eta\,\delta\SN}{6\times10^{-4}}\right)^{1/3}
		\left(\frac{\Sigma\SFRs}{1\,M_\odot\p^{-2}\Gyr^{-1}}\right)^{1/3}
		\left(\frac{n\h}{10^{-3}\cm^{-3}}\right)^{-1/3}		
		\left(\frac{E\SN}{10^{51}\erg}\right)^{1/3}\,,
\end{align}
and, from Eq.~\eqref{Ru},
\begin{align}
U_z\simeq&0.3\,\frac{\!\km}{\!\s} 
		\left(\frac{V_z}{90\kms}\right)
		\left(\frac{f}{0.1}\right)
		\left(\frac{n\h}{10^{-3}\cm^{-3}}\right)
		\left(\frac{h}{0.5\kpc}\right)^{-1}
		\left(\frac{\Sigma\I}{1\,M_\odot\p^{-2}\Gyr^{-1}}\right)^{-1}.
\end{align}

%--------------------------------------------------------------
\subsection{The break-out of a superbubble}\label{OS-BSB}
An alternative estimate is based on a model of a hot, expanding superbubble 
associated with an OB association \citep{MLMC88}. Using an idealized numerical model 
and analytical estimates, these authors argue that a superbubble breaks out of the galactic 
disc when its radius in the plane of about two scale heights of the neutral hydrogen layer. 
The break-out radius is defines as that where the superbubble expansion starts accelerating.
Following \citet{MLMC88} in treating a superbubble as as a large stellar wind bubble, we 
use the expansion law obtained by \citet{WMCCSM77},
\[
r=\left(\frac{125}{154\pi}\right)^{1/5} \left(\frac{Lt^3}{\rho}\right)^{1/5}\,,
\]
where $r$ is the superbubble radius, 
$L=\epsilon\SN\, \eta\, \nu\SN E\SN$ is the mechanical luminosity provided by the supernovae
within an OB association, $t$ is time and $\rho$ is the ambient diffuse gas density. 
We identify the outflow speed with the superbubble expansion speed when $r=2h$ to obtain, 
using the estimate of the supernova rate \eqref{SNrate},
\begin{align}\label{VzML}
V_z\simeq &\left.\frac{\dd r}{\dd t}\right|_{r=2h} 
	= \frac35 \left(\frac{125}{154\pi}\right)^{1/3} 
	  \left(\frac{L}{h\Sigma\I}\right)^{1/3}\nonumber\\
	=&100\,\frac{\!\km}{\!\s}
		\left(\frac{\epsilon\SN\,\eta\,\delta\SN}{6\times10^{-4}}\right)^{1/3}
		\left(\frac{\Sigma\SFRs}{1\,M_\odot\p^{-2}\Gyr^{-1}}\right)^{1/3}		
		\left(\frac{\Sigma\I}{1\,M_\odot\p^{-2}}\right)^{-1/3}\nonumber\\
	&\times\left(\frac{E\SN}{10^{51}\erg}\right)^{1/3}
		\left(\frac{h}{0.5\kpc}\right)^{-1/3}
		\left(\frac{R}{15\kpc}\right)^{2/3}\,,	  
\end{align}
and then 
\begin{align}
U_z\simeq&0.3\,\frac{\!\km}{\!\s} 
		\left(\frac{V_z}{100\kms}\right)
		\left(\frac{f}{0.1}\right)
		\left(\frac{n\h}{10^{-3}\cm^{-3}}\right)
		\left(\frac{h}{0.5\kpc}\right)^{-1}
		\left(\frac{\Sigma\I}{1\,M_\odot\p^{-2}\Gyr^{-1}}\right)^{-1}.
\end{align}

Equations \eqref{VzE} and \eqref{VzML} yield practically identical magnitudes of the
outflow speed and the same relation to the star formation rate but involve different,
if not unrelated, galactic parameters.

For comparison, \citet{ACBMV-M14} find that the maximum velocity of ionized gas outflows 
in a sample of luminous and ultra-luminous infrared galaxies (at low redshifts) scales with 
the star formation rate (SFR) as $V\propto \text{SFR}^a$ with $a=0.24\pm0.05$ for the SFR 
derived from infrared luminosity and $a=0.11\pm0.04$ for the SFR obtained from an 
extinction-corrected H$\alpha$ luminosity. The dependence on the H$\alpha$-derived star 
formation density is similar to that on the corresponding SFR, $V\propto\Sigma_*^{0.13\pm0.03}$. 
\citet{RVS05b} find similar results for the outflow speeds of the neutral gas in
ultra-luminous infrared galaxies (at redshifts 0--0.5) and four dwarf starburst galaxies, 
with $a=0.24\pm0.04$ for SFR derived from the infrared luminosity. The variation of $V$ with
SFR apparently flattens at $\text{SFR}\gtrsim10\,M_\odot\yr^{-1}$ \citep{M05,RVS05b,ACBMV-M14,Metal12}. 
The samples of both \citet{ACBMV-M14} and \citet{RVS05b} mainly contain such galaxies. 
\citet{M05} considered galaxies at redshifts $0.042$-–$0.16$, some of which
have lower SFRs, to obtain $V\propto\text{SFR}^{0.35\pm0.06}$ for the upper
envelope of the data points in the $(V,\text{SFR})$ plane \citep[see also][]{Metal12}.
This dependence agrees very well with that in Eqs~\eqref{VzE} and \eqref{VzML}, 
provided $\text{SFR}\propto\Sigma_*$.

\clearpage

\end{document}

%% file: table1.tex
%%%
% Table 1:
%%%%

\begin{table*}
\centering
\begin{threeparttable}
\caption{\label{galaxybackground}The general properties of the sample galaxies and the 
method used to estimate the total and mean
magnetic field strengths, $\B$ and $\BB$, given in Table~\ref{tabledata}. 
Blank entries (---) indicate that the
data are not available. References can be found in Table~\ref{tableBrefs1}.
}
\begin{tabular}{lllcccccc}
\hline\hline
Galaxy\phantom{\Big[}%taller line, to avoid overbar over B merging with \hline
				&\multicolumn{2}{c}{Hubble Type\tnote{$\star$}}
                          &    &Distance\tnote{$\dagger$}
                                                         &Linear resolution\tnote{$\S$}
                                                                           &Method\tnote{$\P$}
                                                                                              &\multicolumn{2}{c}{Inclination{\tnote{$\parallel$}}\ \mbox{ [$^\circ$]}}\\
          \cline{2-3}                                                                          \cline{8-9}
         &NED        &LEDA &
                             &[Mpc]
                                                        &[kpc]        &    &REF  &LEDA\\
(1)      &(2)         &(3)   & &(4) & (5)
                                                        &(6) &(7) & (8) \\
\hline
M31      &SA(s)b      & Sb&   &0.69          &0.6/1.0/0.15   &E    &78  &72\\
M33      & SA(s)cd    & Sc  & &0.84           &0.7      &F   &56     &55 \\
M51      & Sa+Sc      & Sbc&  &7.6     &0.6      &F   &20     &33 \\
M66      &SAB(s)b     & SABb& &11.9    &0.8        &E  &---    &68 \\
M81      & SA(s)ab    & Sab & &3.25     &0.7:1.1        &E  &59     &63 \\
M82      & I0         &Scd &  &5.0   &0.58:0.17&E  &---    &77 \\
M94      & (R)SA(r)ab & Sab&  &4.7          &0.3       &E  &35     &32 \\
M99      & SA(s)c     & Sc &  &20.0           &1.5        &E  &42     &20 \\
M104     &SA(s)a      & Sa  & &8.9               &3.6        &E  &84     &59 \\
M109     & SB(rs)bc   & Sbc & &15.0          &2.2        &E  &59     &47 \\
NGC 253  & SAB(s)c    &SABc&  &3.94              &0.6/1.6/2.8        &E  &79     &90 \\
NGC 891  & SA(s)b?    & Sb&   &7.2           &1.4/2.8:2.0        &E  &$\ge88$&90 \\
NGC 1097 & SB(s)b     &SBb &  &17.0           &0.8        &F  &45     &55 \\
NGC 1365 & SB(s)b     & Sb  & &18.6             &2.3        &E  &40     &63 \\
NGC 1566 & SAB(s)bc   & SABb& &17.4           &2.2        &E  &27     &48 \\
NGC 4414 &SA(rs)c?    & Sc &  &19.2            &1.5         &E  &55     &57 \\
NGC 5775 & SBc?       & SBc  &&26.7           &2.1          &E  &86     &83 \\
NGC 5907 & SA(s)c?    &SABc & &11.0           &2.2        &E  &87     &90 \\
NGC 6946 & SAB(rs)cd  & SABc& &5.5 		&0.4   		 &E  	&38     &18 \\
IC 342   & SAB(rs)cd  & SABc& &3.1            &4/2.4        &E  &25     &19 \\
\hline
\end{tabular}
\begin{tablenotes}[flushleft]
\footnotesize
\item[$\star$]
				Hubble type according to the NED (http://ned.ipac.caltech.edu)
				and LEDA (http://leda.univ-lyon1.fr) databases.
\item[$\dagger$]
				Distance to the galaxy, mostly adopted to be same as in the original publication of
				the magnetic field data or, if not given there, of the rotation curve data
				(see Table~\ref{tableBrefs1}).
\item[$\S$]
				The linear resolution of the magnetic field observations: entries separated
				by a solidus are those at the individual observation wavelengths wherever the resolution
				of the data analysis is not specified; those separated by a colon represent the major
				and minor axes of the beam.
\item[$\P$]
				The method used to estimate $\B$ and $\BB$: equipartition with cosmic rays
				and the degree of polarization (E) or the Faraday rotation (F).
\item[$\parallel$]
				The inclination angle, with $90^\circ$ corresponding to the edge-on view:
				entries taken from the sources shown in Table~\ref{tableBrefs1} are in Column~(7),
				and those form the LEDA database, in Column~(8).
\end{tablenotes}
\end{threeparttable}
\end{table*}

%%%

%% file: table2.tex
%%%%%%%
%
% Table 2
%
%%%%%%%

\begin{table*}
\centering
\begin{threeparttable}
\caption{\label{tabledata}
The magnetic field and ISM parameters, in the radial ranges specified; the
ranges marked with an asterisk are not given explicitly in the original
publication (Table \ref{tableBrefs1}), and have been estimated from the
extent of polarized emission. A negative value of $p_B$ corresponds to a
trailing spiral. Blank entries appear wherever the data could
not be found, or were not found in a usable form.
}
\begin{tabular}{lr@{--}lcccccccccr}
\hline\hline
Galaxy\phantom{\Big[}
               &\multicolumn{2}{c}{Radial range}  &$\B$ &$\BB$ &$p_B$   &$\BB_0$ & $\Sigma_\mathrm{I}$ &$\Sigma_2$ &$\Sigma\SFRs$ &$\Omega$ &$S$ &$\overline{\Omega S}$\phantom{000}\\ 
\phantom{\Bigg[}               &\multicolumn{2}{c}{[kpc]}         &	[$\mu$G]	 &    [$\mu$G]     &[$^\circ$]   &[$\mu$G] &$\left[\dfrac{M_\odot}{\!\p^2}\right]$
                                					&$\left[\dfrac{M_\odot}{\!\p^2}\right]$
                                							  &$\left[\dfrac{M_\odot}{\!\p^2\Gyr}\right]$
                                                                  		&$\left[\dfrac{\rm km}{\rm s\,kpc}\right]$
                                                                  				&$\left[\dfrac{\rm km}{\rm s\,kpc}\right]$ 
                                                                  					& $\left[\dfrac{\rm km^2}{\rm s^2\,kpc^2}\right]$ \\
\hline
M31	&	6	&	8	&	7.3	&	4.9	&	$-$13	&	4.8	&	1.47	&	0.266	&	0.443	&	38.4	&	$-$28.7	&	$-$1100	\\
	&	8	&	10	&	7.5	&	5.2	&	$-$19	&	5.6	&	2.17	&	0.308	&	0.621	&	31.1	&	$-$30.9	&	$-$959	\\
	&	10	&	12	&	7.1	&	4.9	&	$-$11	&	4.7	&	3.64	&	0.665	&	0.794	&	25.1	&	$-$26.9	&	$-$680	\\
	&	12	&	14	&	6.3	&	4.6	&	$-$8	&	4.9	&	4.05	&	0.51	&	0.227	&	21.1	&	$-$21.6	&	$-$458	\\
M33	&	0.25	&	6.75	&	6.1	&		&		&		&	8.86	&	1.22	&	4.05	&	25.3	&	$-$18.6	&	$-$507	\\
	&	1	&	3	&		&	1.4	&	$-$48	&	0.7	&	11.3	&	1.90	&	9.64	&	40.7	&	$-$25.4	&	$-$1050	\\
	&	3	&	5	&		&	0.6	&	$-$42	&	0.3	&	9.43	&	1.28	&	3.99	&	24.9	&	$-$20.8	&	$-$523	\\
M51	&	2.4	&	7.2	&	20	&		&		&		&	6.93	&	16.8	&	14.0	&	52.5	&	$-$53.7	&	$-$3160	\\
	&	2.4	&	3.6	&		&	6.2	&	$-$20	&	1.2	&	5.20	&	27.1	&	20.7	&	86.5	&	$-$97.4	&	$-$8380	\\
	&	3.6	&	4.8	&		&	6.8	&	$-$24	&	1.5	&	5.97	&	19.3	&	13.5	&	58.1	&	$-$60.9	&	$-$3640	\\
	&	4.8	&	6	&		&	4.5	&	$-$22	&	2.5	&	8.98	&	14.9	&	18.0	&	46.7	&	$-$40.6	&	$-$1890	\\
	&	6	&	7.2	&		&	3.5	&	$-$18	&	2.5	&	6.63	&		&	7.92	&	38.1	&	$-$40.0	&	$-$1540	\\
M66	&	0	&	7*	&	11	&	4	&		&		&	5.16	&	19.5	&	13.7	&	41.8	&	$-$37.0	&	$-$1590	\\
M81	&	3	&	6	&	7.5	&	3	&		&		&	2.62	&		&	0.954	&	51.0	&	$-$42.7	&	$-$2250	\\
	&	6	&	9	&		&		&	$-$21	&		&	3.33	&		&		&	31.7	&	$-$39.2	&	$-$1240	\\
	&	9	&	12	&		&		&	$-$26	&		&	2.32	&		&		&	19.8	&	$-$29.5	&	$-$599	\\
M82	&	0	&	0.5	&	50	&		&		&		&		&	430	&	3000	&		&		&		\\
M94	&	0	&	2.4*	&	17	&	10	&	$-$35	&		&	5.38	&	16.2	&	22.4	&	130	&	$-$156	&	$-$23800	\\
M99	&	0	&	11.5*	&	16	&	7	&		&		&	7.91	&	11.0	&	14.2	&	28.3	&	$-$25.8	&	$-$863	\\
M104	&	0	&	5*	&	6	&	3	&		&		&		&		&		&	95.1	&	$-$71.7	&	$-$14800	\\
M109	&	0	&	17.5*	&	6	&	2	&		&		&	2.96	&		&		&	24.7	&	$-$20.0	&	$-$503	\\
NGC253	&	0	&	8*	&	15	&	4.4	&	$-$25	&		&	3.28	&	2.35	&	35.1	&	50.9	&	$-$49.2	&	$-$5390	\\
NGC891	&	0	&	7.7*	&	13	&	6.9	&		&		&	3.28	&	17.0	&	10.4	&	55.5	&	$-$44.9	&	$-$7060	\\
NGC1097	&	0	&	3.75*	&	13	&	3	&		&		&	2.94	&	52.7	&	631	&	158	&	$-$167	&	$-$22100	\\
	&	1.25	&	2.5	&		&		&	$-$34	&	1.4	&	3.00	&	60.2	&		&	182	&	$-$219	&	$-$41800	\\
	&	2.5	&	3.75	&		&		&	$-$36	&	1.1	&	2.95	&	44.7	&		&	94.5	&	$-$135	&	$-$13100	\\
	&	3.75	&	5	&		&		&	$-$23	&	1.9	&	3.13	&	30.0	&		&	62.1	&	$-$61.8	&	$-$3940	\\
NGC1365	&	0	&	14*	&	9	&	5.7	&		&		&	8.16	&		&		&	36.6	&	$-$38.5	&	$-$3340	\\
	&	2.625	&	4.375	&		&		&	$-$34	&	0.8	&	3.39	&		&		&	71.3	&	$-$41.7	&	$-$2990	\\
	&	4.375	&	6.125	&		&		&	$-$17	&	0.8	&	4.31	&		&		&	52.4	&	$-$50.0	&	$-$2630	\\
	&	6.125	&	7.875	&		&		&	$-$31	&	0.7	&	6.89	&		&		&	39.3	&	$-$41.1	&	$-$1620	\\
	&	7.875	&	9.625	&		&		&	$-$22	&	1.2	&	9.50	&		&		&	30.9	&	$-$34.2	&	$-$1060	\\
	&	9.625	&	11.375	&		&		&	$-$37	&	1.1	&	10.8	&		&		&	25.1	&	$-$29.8	&	$-$750	\\
	&	11.375	&	13.125	&		&		&	$-$29	&	0.7	&	9.58	&		&		&	20.9	&	$-$25.4	&	$-$530	\\
	&	13.125	&	14.875	&		&		&	$-$33	&	0.4	&	8.26	&		&		&	17.7	&	$-$21.4	&	$-$380	\\
NGC1566	&	0	&	10*	&	13	&	3	&		&		&	9.18	&		&		&	33.4	&	$-$32.3	&	$-$1180	\\
	&	2.7	&	8	&	8	&		&	$-$20	&		&	9.69	&		&		&	38.2	&	$-$36.6	&	$-$1560	\\
NGC4414	&	0	&	5.4*	&	15	&	4	&	$-$22	&		&	9.32	&	90.5	&	30.6	&	79.5	&	$-$70.4	&	$-$7240	\\
NGC5775	&	0	&	13.5*	&	11	&	4	&		&		&		&		&		&	25.2	&	$-$23.7	&	$-$976	\\
NGC5907	&	0	&	8*	&	5	&	1.1	&		&		&		&		&	5.13	&	54.7	&	$-$46.4	&	$-$3920	\\
NGC6946	&	0	&	9.2*	&	14.7	&	6.3	&		&		&	6.00	&	10.2	&	11.9	&	34.0	&	$-$31.7	&	$-$2070	\\
	&	0	&	6	&		&		&	$-$27	&		&	6.30	&	22.2	&	20.2	&	48.1	&	$-$41.2	&	$-$4090	\\
	&	6	&	12	&		&		&	$-$21	&		&	4.08	&	0.588	&	2.50	&	19.6	&	$-$20.5	&	$-$422	\\
	&	12	&	14	&		&		&	$-$10	&		&		&		&		&	13.4	&	$-$13.7	&	$-$184	\\
IC342	&	0	&	13.5	&	9	&	4	&		&		&	6.13	&	5.76	&	16.3	&	27.8	&	$-$27.0	&	$-$1480	\\
	&	5	&	9	&		&		&	$-$21	&		&	6.41	&	6.16	&		&	28.1	&	$-$29.4	&	$-$841	\\
	&	9	&	13	&		&		&	$-$26	&		&	6.53	&		&		&	17.7	&	$-$16.9	&	$-$301	\\

\hline
\end{tabular}
\begin{tablenotes}[flushleft]
\footnotesize
\item[]
The magnetic field pitch angles shown are either azimuthally averaged values or,
where observations have been interpreted in finer detail,
are the values for the axisymmetric component of the magnetic field.
$\BB_0$ is the strength of the axisymmetric component of the mean magnetic field.
\end{tablenotes}
\end{threeparttable}
\end{table*}

%% file: table3.tex
%%%%%%%%%%
%
% Table 3:  Van Eck et al. 2014
%
%%%%%%%%%%

\begin{table*}
\caption{\label{tableBrefs1}
Data sources.}
\centering
\begin{tabular}{@{}lllll@{}}
\hline \hline
Galaxy\phantom{\Big[}
					&$\B$, $\BB$, $\BB_0$ and inclination 	 &Rotation curve      &Gas density           &Star formation rate  \\
\hline
M31 (NGC 224)  &\citet{Fletcher04}       &\citet{Sofue99}     &\citet{Boissier07}    &\citet{Tabatabaei10} \\
M33 (NGC 598)  &\citet{Tabatabaei08}     &\citet{Sofue99}     &\citet{Boissier07}    &\citet{Verley09}     \\
M51 (NGC 5194) &\citet{Fletcher11}       &\citet{Sofue99}     &\citet{Leroy08}       &\citet{Leroy08}      \\
M66 (NGC 3627) &\citet{Soida01}          &\citet{deBlok08}    &\citet{Leroy08}       &\citet{Leroy08}      \\
M81 (NGC 3031) &\citet{Krause89}         &\citet{Sofue99}     &\citet{Boissier07}    &\citet{Calzetti10}   \\
M82 (NGC 3034) &\citet{Klein88}          &\citet{Sofue99}     &\citet{Lo87}          &\citet{Lo87}         \\
M94 (NGC 4736) &\citet{Chyzy08-4736}     &\citet{deBlok08}    &\citet{Crosthwaite01} &\citet{Leroy08}      \\
M99 (NGC 4254) &\citet{Chyzy08}          &\citet{Dicaire08}   &\citet{Warmels88}     &\citet{Rahman11}     \\
M104 (NGC 4594)&\citet{Krause06}         &\citet{Tempel06}    &---                   &--- 								 \\
M109 (NGC 3992)&\citet{Beck02}           &\citet{Bottema02}   &\citet{Bottema02}     &--- 								 \\
NGC 253        &\citet{Heesen09}         &\citet{Sofue99}     &\citet{Sorai00}       &\citet{Waller88}     \\
NGC 891        &\citet{Hummel91}         &\citet{Yim11}       &\citet{Yim11}         &\citet{Yim11}        \\
NGC 1097       &\citet{Beck05}           &\citet{Sofue99}     &\citet{Crosthwaite01} &\citet{Kennicutt98}  \\
NGC 1365       &\citet{Beck05}           &\citet{Sofue99}     &\citet{Jalocha10}     &---                  \\
NGC 1566       &\citet{Ehle96}           &\citet{Sofue99}     &\citet{Becker88}      &---                  \\
NGC 4414       &\citet{Soida02}          &\citet{Fridman05}   &\citet{Thornley97}    &\citet{Wong02}       \\
NGC 5775       &\citet{Soida11}          &\citet{Heald06}     &---                   &---                  \\
NGC 5907       &\citet{Dumke00}          &\citet{Brownstein06}&---                   &\citet{Misiriotis01} \\
NGC 6946       &\citet{Beck07,EB93}      &\citet{deBlok08}    &\citet{Leroy08}       &\citet{Leroy08}      \\
IC 342         &\citet{Graeve88}         &\citet{Sofue99}     &\citet{Crosthwaite01} &\citet{Calzetti10}   \\
\hline
\end{tabular}
\end{table*}

%% file: table4.tex
%%%%%%
% TABLE 5 - Van Eck et al. 2014
%%%%%%

\begin{table*}
\caption{\label{tablestrengthcorrelations}
Kendall's correlation coefficient $\tau$, the significance level $\nu$ of the correlation, 
and the number of galaxies (or galaxy regions) involved $N$ for the inter-dependencies between the 
large-scale magnetic field strength obtained from observations and from various 
models of Section~\ref{RMFS}. Entries shown in bold pass the 5\% significance level test.}
\centering
\begin{tabular}{lccccccc}
\hline\hline
Model 				    &\multicolumn{3}{c}{$\BB_0^2$} & &\multicolumn{3}{c}{$\tan p_B$} \\     %alternative results for regular field strength \BB:
                             \cline{2-4}                      \cline{6-8}
                             &$\tau$  &$\nu$, \% &$N$          & &$\tau$  &$\nu$, \% 	&$N$\\
\hline
Energy equipartition, 
Eq.~\eqref{energy_bal_obs}   &$-$0.005 & 97 & 20			     & &&	&\\  % 0.15 &31 &23
Kinematic, Eq.~\eqref{pBkn} & & & & 								&0.013 &92 & 31 \\ 
Magnetostrophic balance, 
Eqs~\eqref{strophic_bal_obs} and \eqref{pitch_strophic_obs} &$-$0.091 & 57 & 20        	   & &$-$0.10 &62 	&15\\ % 0.34 &5.6 &17
Magnetic helicity balance, 
Eqs~\eqref{hel} and \eqref{pitch_z} 	 &$-$0.405 & 10 & 10           & 			&$-$0.10 &62 	&15\\  % 0.28 &11 &17
\hline
\end{tabular}
\end{table*}

%% file: table5.tex
%%%%
% Table 4 - Van Eck et al. 2014
%%%%

\begin{table*}
\caption{\label{tabletaucoeffs}
Kendall's correlation coefficient $\tau$, the significance level $\nu$ and the
number of galaxies (or galaxy regions)  involved $N$ in the pair-wise correlations of magnetic field
parameters and other characteristics of the interstellar medium. Entries shown in 
bold pass the $\nu=5$\% significance level test. For the significant correlations,
we also present the best-fit power-law index $k$ as, e.g., in $\B\propto\Sigma_2^{0.24\pm0.04}$.
}
%\centering
\begin{tabular}{lccccccccccccccc}
\cline{1-15}
\phantom{$\displaystyle\int$}					    
														 &\multicolumn{4}{c}{$\B$}        & &\multicolumn{4}{c}{$\BB$}& & &\multicolumn{4}{c}{$\tan p_B$} \\
                             \cline{2-5}                       \cline{7-10}                    \cline{12-15}
                             &$\tau$		&$\nu$, \%	&$N$ 	&$k$& &$\tau$	&$\nu$, \%&$N$		&$k$ &&$\tau$	&$\nu$, \% 	&$N$ &$k$\\
\cline{1-15}
$\B$ 												 & 					& 					& 		& & &$\bf\phz0.42$   	&\bf1.0 	&20 &$0.76\pm0.23$ \\
$\Sigma_\mathrm{I}$          &$\phz0.26$    	&13\phzz  &19 &              & &$\phz-0.14$ &36 &23     &  & &$-0.21$ 		&$10\phzz$	&30 	&\\
$\Sigma_2$                   &$\phz\bf0.50$ 	&\bf0.6 	&16 &$0.21\pm0.04$ & &$\phz0.02$ &91 &18     &  & &$\bf-0.36$ 	&$\bf4.3$ 	&17 	&$-0.10\pm0.08$\\
$\Sigma_\mathrm{I}+\Sigma_2$ &$\bf\phz0.43$    	&\bf2.5\phzz  &15 & $0.24\pm0.07$             & &$-0.03$ &85 &18      & & &$\bf-0.39$ 	&$\bf2.8$ 	&17 	&$-0.25\pm0.13$\\
$\Sigma\SFRs$                &$\phz\bf0.54$ 	&\bf0.17 	&18 &$0.19\pm0.03$ & &$\phz0.10$ &52 &21      & & &$\bf-0.44$ 	&$\bf2.2$ 	&15 	&$-0.15\pm0.09$\\
$\Omega$                     &$\phz0.29$    	&6.2\phzz  &22 	&             & &$\phz0.06$ &65 &26   &   & &$-0.16$ 		&$21\phzz$  	&31 \\
$-S$                         &$\bf0.34$       	&\bf2.7\phzz  &22 	&             & &$0.13$    &34 &26    &  & &$\phz-0.14$ 	&$\phz26\phzz$  &31 \\
$-\overline{\Omega S}$       &$\bf0.34$       	&\bf2.3\phzz  &22 	& $0.14\pm0.04$            & &$0.13$    &34 &26     & & &$\phz-0.16$ 	&$\phz22\phzz$  &31 \\
$\BB_0\phantom{\displaystyle\int}$ & & & & & & & & &  & &$\bf\phz0.56$    	&\bf0.048 	&20 & $0.51\pm0.11$ \\
\cline{1-15}
\end{tabular}
\end{table*}